\documentclass{article}

\usepackage{rotating}
\usepackage{latexsym}
\usepackage{graphics}
\def\nue{$\nu_e$}
\def\numu{$\nu_\mu$}
\def\numunue{$\nu_\mu \rightarrow \nu_e$}

\newcommand{\sizedfig}[2]{\resizebox{#1}{!}{\includegraphics{#2}}}

\newlength{\overlaywd}
\newlength{\overlayht}

\begin{document}

\title{Event Rates for Off Axis NuMI Experiments}
\author{B. Viren, bv@bnl.gov}
\maketitle
\begin{abstract}
  Neutrino interaction rates for experiments placed off axis in the
  NuMI beam are calculated.  Primary proton beam energy is 120 GeV and
  four locations at 810 km from target and 6, 12, 30 and 40 km off
  axis are considered.  This report is part of the Joint FNAL/BNL
  Future Long Baseline Neutrino Oscillation Experiment Study.
\end{abstract}

\tableofcontents
\clearpage

\section{Introduction}

A generic calculation of neutrino event rates for detectors at various
locations in the NuMI neutrino beam has been done.  Only flux,
probability, cross sections and rudimentary energy reconstruction is
considered.  No particular detector technology is assumed.

\section{Baselines}

This document gives a calculation of neutrino flux and interaction
rates for detectors placed in the various locations in the NuMI beam
at 810 km from the target.  Four off-axis locations are considered: 6
km (7.4 mrad), 12 km (14.8 mrad), 30 km (37.0 mrad) and 40 km (49.4
mrad).

\section{Neutrino Flux}

Neutrino flux calculations are performed using the GEANT3 based GNUMI
simulation program.  The proton beam, target, focussing horns, decay
tunnel and other elements are models of what is currently in use by
MINOS.  The locations of the two focusing horns with respect to the
target can focus the neutrino parents to produce different spectra.
The so called ``Low Energy'' (LE)\footnote{More properly, MINOS calls this ``LE-10''} and ``pseudo Medium Energy'' (pME)
tuning are used to confirm the simulation against measured MINOS near
detector data while the ``Medium Energy'' tuning is used for spectra
at 810 km to match.  Table~\ref{tab:beams} summarizes the beam tunings
used.  In all cases a 120 GeV primary proton beam is used.

\begin{table}[htbp]
  \centering
  \begin{tabular}{|r|r|r|r|r|}
    \hline
    Name & Target (cm) & Horn 2 (m) & Current (kA) & Version\\
    \hline
    LE   & -10              & +10              & 182 & v18 \\
    pME  & -100             & +10              & 197 & v18 \\
    ME   & -100             & +13              & 182 & v15 \\
    \hline
  \end{tabular}
  \caption{Summary of Neutrino Flux Spectra.  Distances are measured w.r.t. face of horn 1.}
  \label{tab:beams}
\end{table}

\begin{figure}[htbp]
  \centering
  \sizedfig{\textwidth}{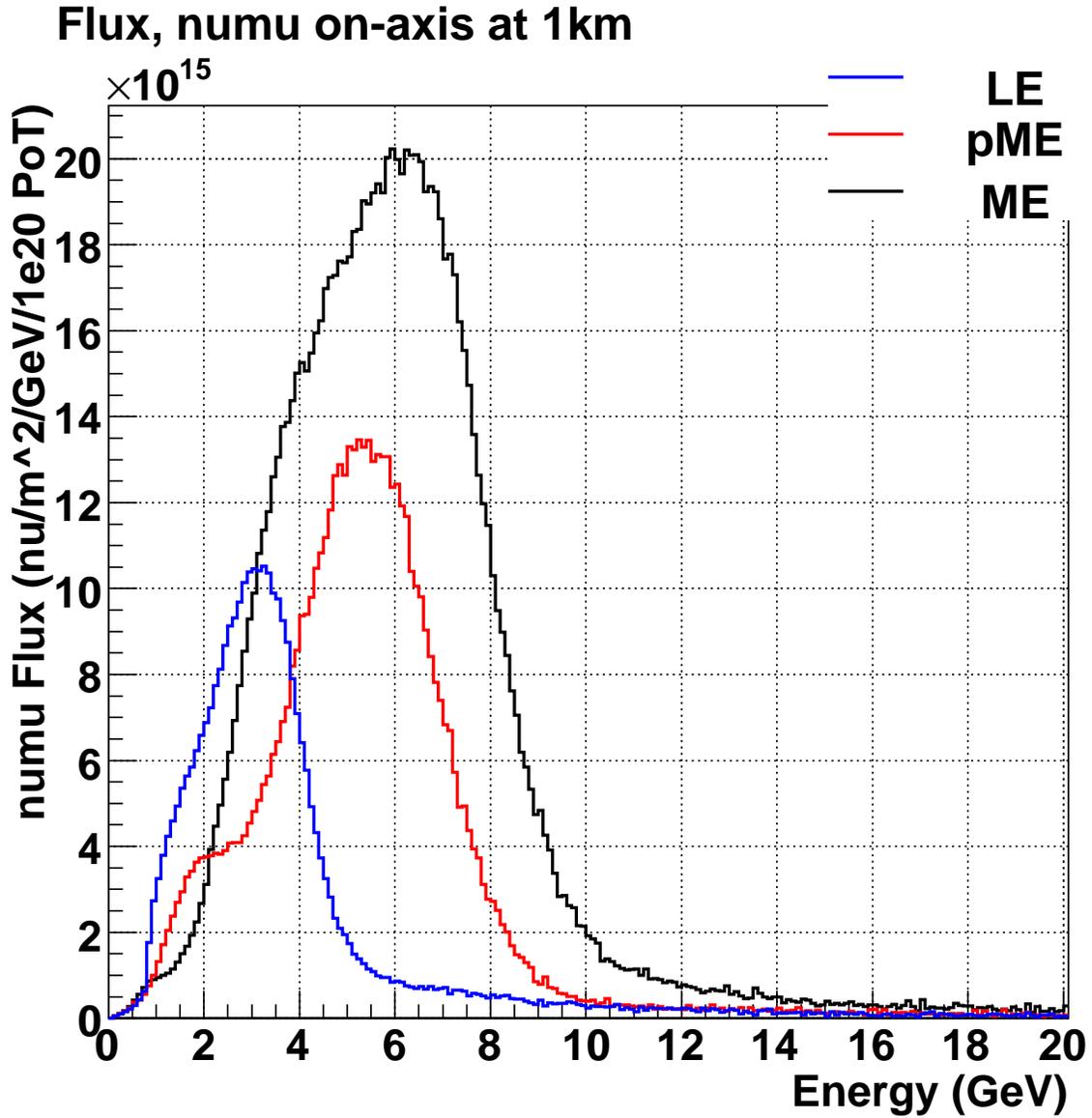}  
  \caption{Flux spectra of \numu{} neutrinos at 1km for the beams considered.  Note, this is not far flux scaled to 1km.  It contains effects of the secondary beam being an extended source.}
  \label{fig:flux-at-1km-numu}
\end{figure}

\begin{figure}[htbp]
  \centering
  \sizedfig{\textwidth}{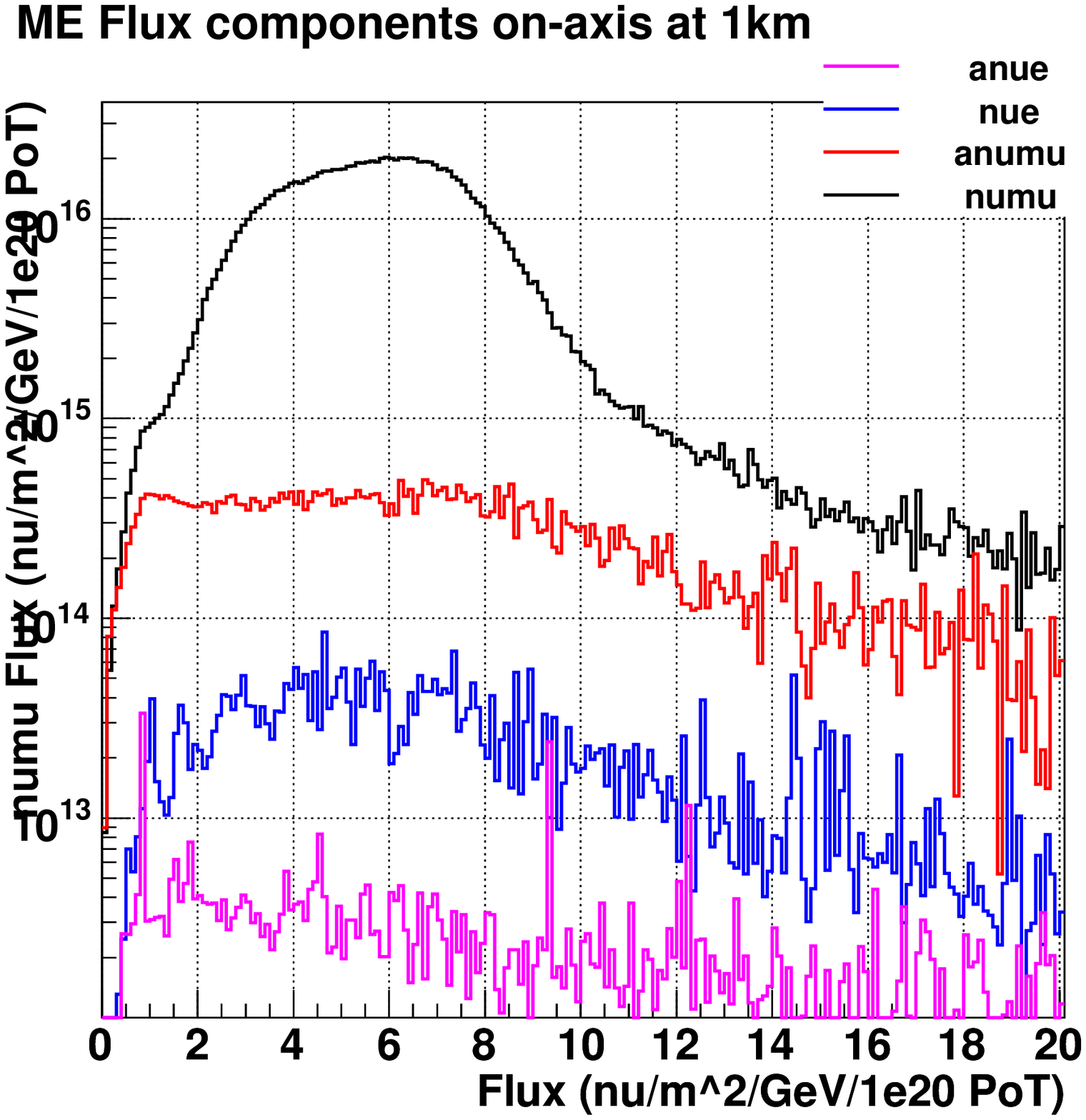}
  \caption{Neutrino flux spectra at 1km for the ME beam.}
  \label{fig:flux-at-1km-me}
\end{figure}

Figure~\ref{fig:flux-at-1km-numu} shows the \numu{} spectra for these
three beam configurations.  Figure~\ref{fig:flux-at-1km-me} shows the
four neutrino components of the flux for the ME configuration.

\section{Oscillation Probability}

\begin{figure}[htbp]
  \centering
  \sizedfig{0.49\textwidth}{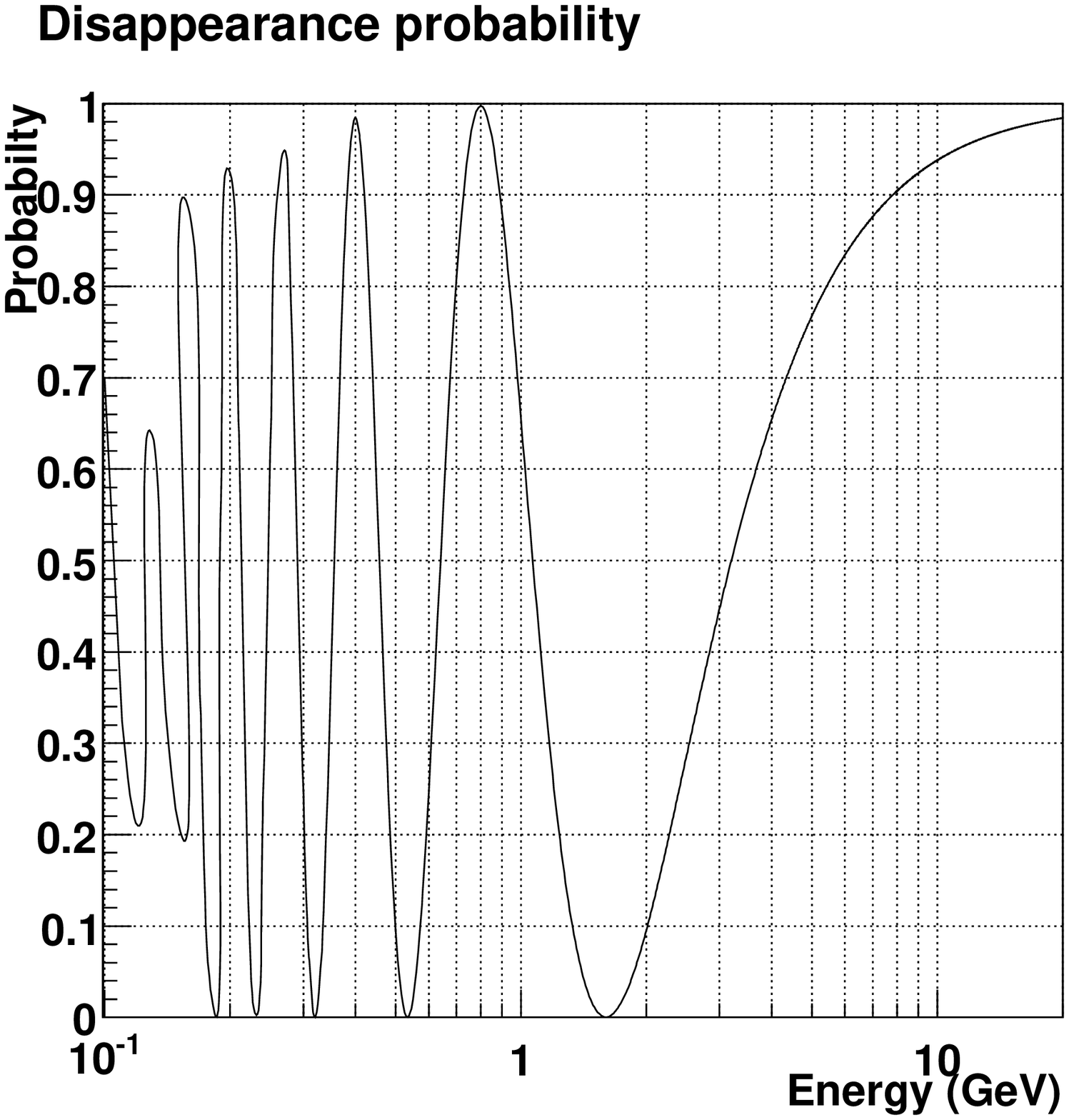}%
  \sizedfig{0.49\textwidth}{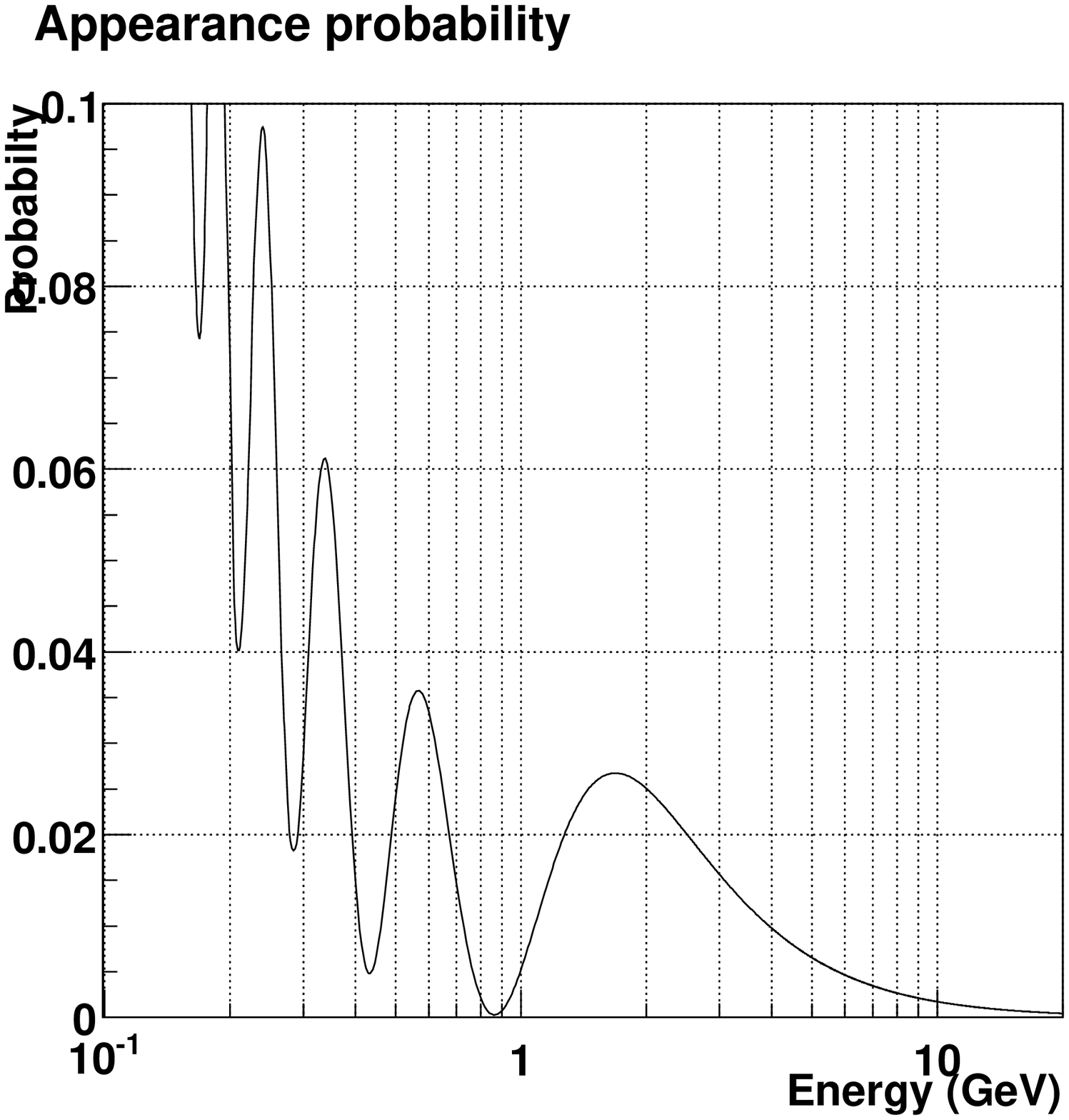}%
  \caption{Disappearance (left) and appearance (right) neutrino oscillation probabilities as a function of the neutrino energy.  See table~\ref{tab:prob} for oscillation parameters.}
  \label{fig:prob}
\end{figure}

For this study, the neutrino oscillation parameters that were used are
given in Table~\ref{tab:prob}.  They were calculated using the program
``nuosc'' part of libnuosc++\cite{nuosc}.  This calculation is a full
three-neutrino numerical calculation using the Preliminary Reference
Earth Model (PREM)\cite{prem} for Earth's density.  The calculation
mode using constant matter densities averaged over the baseline was
used.  Figure~\ref{fig:prob} shows the disappearance and appearance
probabilities used.

\begin{table}[htbp]
  \centering
  \begin{tabular}{|r|r|}
    \hline
    Parameter & Value \\
    \hline
    $\Delta m^2_{31}$ & $2.5\times 10^{-3}eV^2$ \\
    $\Delta m^2_{21}$ & $8.6\times 10^{-5}eV^2$ \\
    $\sin^22\theta_{12}$ & $0.86$ \\
    $\sin^22\theta_{23}$ & $1.0$ \\
    $\sin^22\theta_{13}$ & $0.04$ \\
    $\delta_{CP}$ & $0$ \\
    \hline    
  \end{tabular}
  \caption{Neutrino oscillation parameters used in this study.}
  \label{tab:prob}
\end{table}

\section{Cross Sections}

The interactions considered are:

\begin{itemize}
\item Quasi-elastic (QE) charged current (CC)
\item Total charged current
\item Neutral current (NC) single $\pi^0$ productions (1$\pi^0$)
\end{itemize}

\begin{figure}[htbp]
  \centering
  \sizedfig{\textwidth}{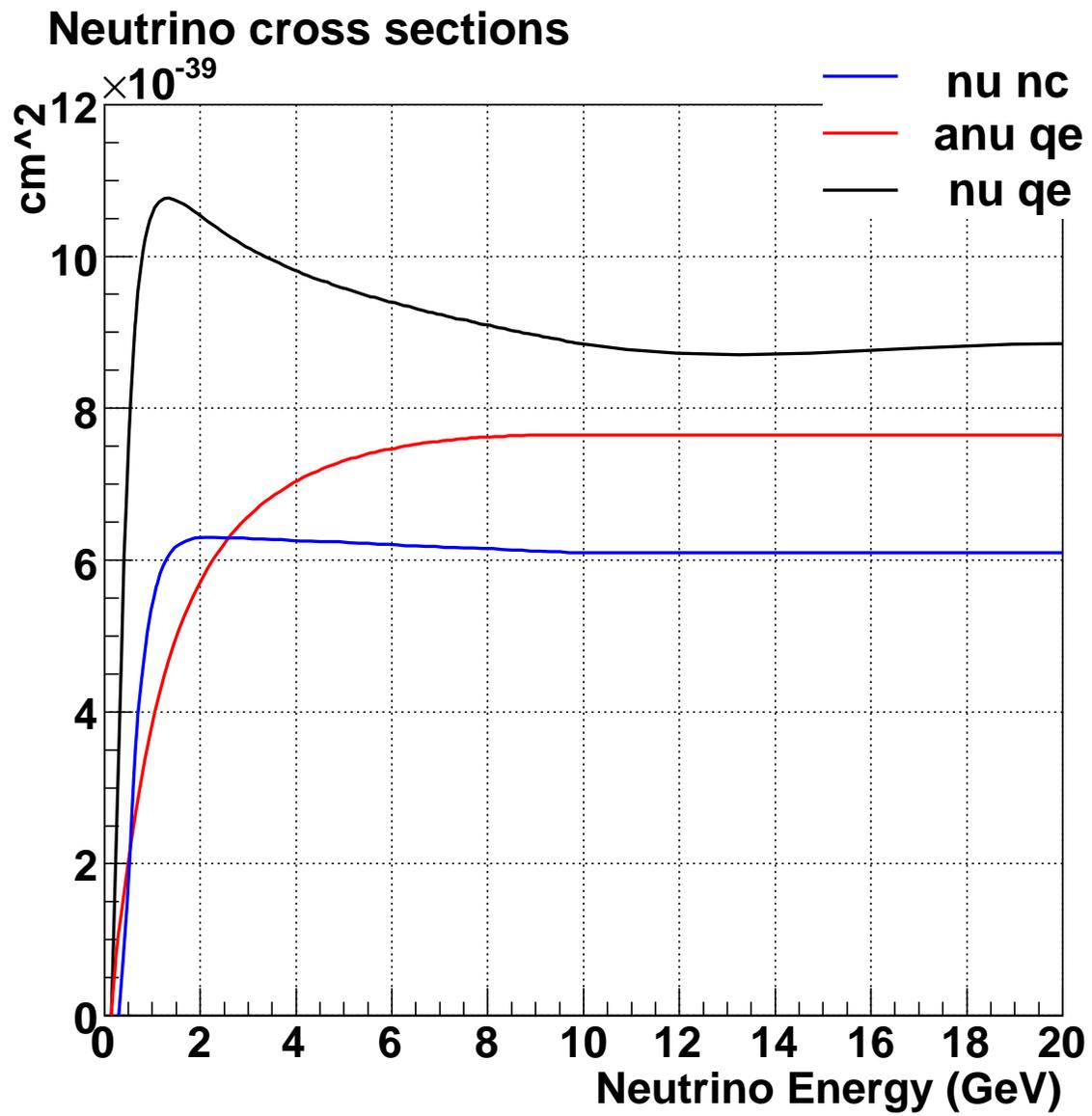}
  \caption{QE and NC-1$\pi^0$ cross sections.}
  \label{fig:cross-section}
\end{figure}

The cross sections used for QE and NC-1$\pi^0$ are shown in
Figure~\ref{fig:cross-section}.  The total CC cross sections are
parameterized for neutrinos as:

\[\sigma_{\nu_\mu,CC} = 0.80\times 10^{-38} cm^2/GeV \times E_\nu\]

and anti-neutrinos as:

\[\sigma_{\bar\nu_\mu,CC} = 0.35\times 10^{-38} cm^2/GeV \times E_\nu\]

The neutrino energy in charged current (total and QE) events is
assumed to be reconstructed with perfect energy resolution and with no
systematic bias.  For NC-1$\pi^0$ events the reconstructed neutrino
energy is taken to be the true energy of the $\pi^0$ and no
consideration for shower angle w.r.t. to the incoming neutrino is
made.  This produces a softer reconstructed energy spectrum than would
be found if this angle were considered.  In addition, no
account of nuclear absorption nor charge exchange by the $\pi^0$ is
made.

\begin{figure}[htbp]
  \centering
  \sizedfig{0.8\textwidth}{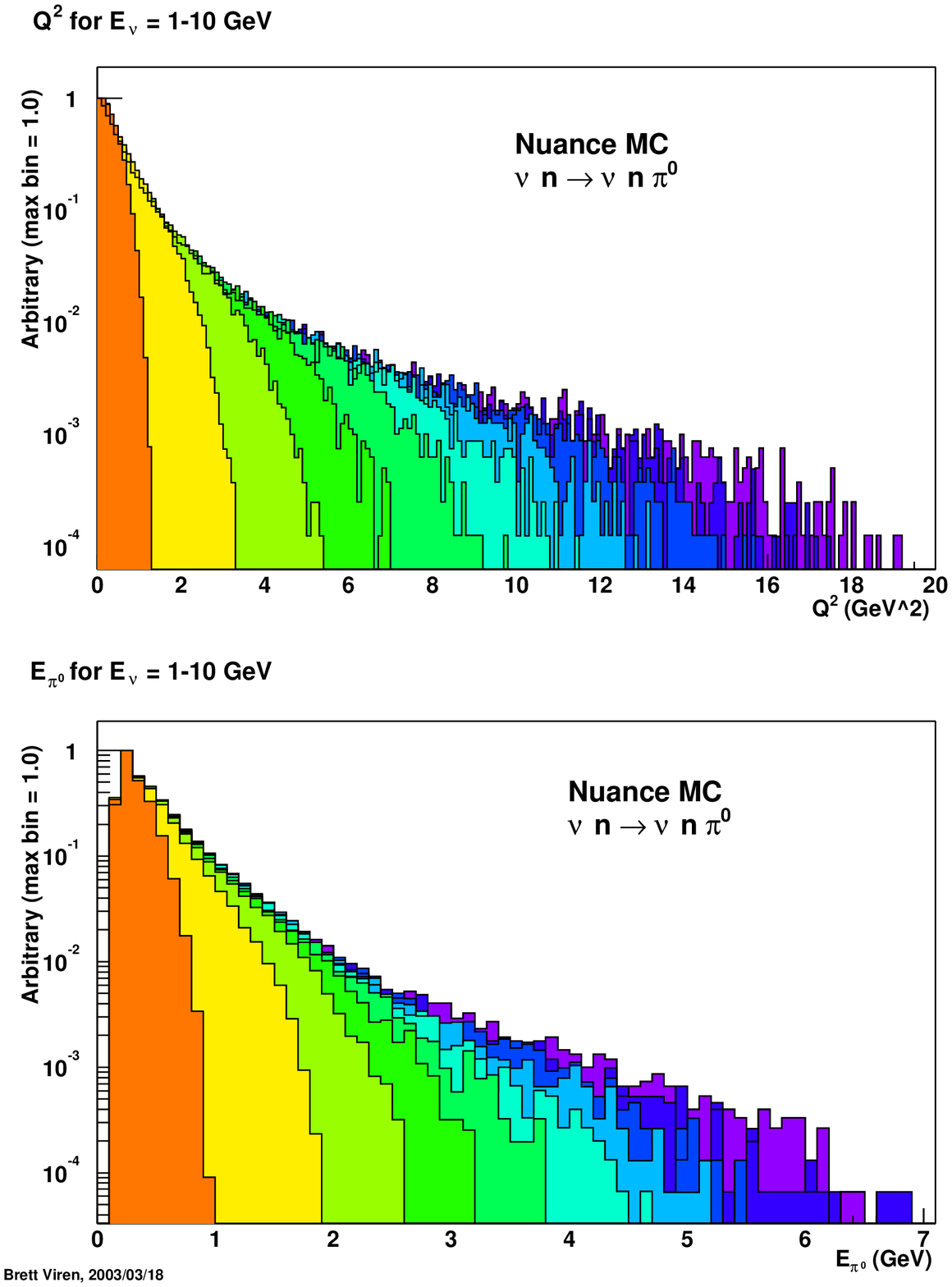}
  \caption{(a) $q^2$ (top) and (b) $\pi^0$ energy (bottom) as functions of mono-energetic neutrinos in 1 GeV steps.  Calculation uses 0.5 GeV steps in neutrino energy.}
  \label{fig:epi0}
\end{figure}

To simulate the energy of the $\pi^0$ sets of $E_{\pi^0}$ spectra were
generated using mono-energetic neutrinos with energies chosen in steps
of 0.5 GeV.  These were generated with the NUANCE\cite{nuance} simulation.
Figure~\ref{fig:epi0} illustrates this.

All cross sections are applied assuming the detector mass is made up
of equal numbers of protons and neutrons.

\section{Comparison with MINOS Near Detector Data}

\begin{sidewaysfigure}[htbp]
  \centering
  \sizedfig{0.33\textwidth}{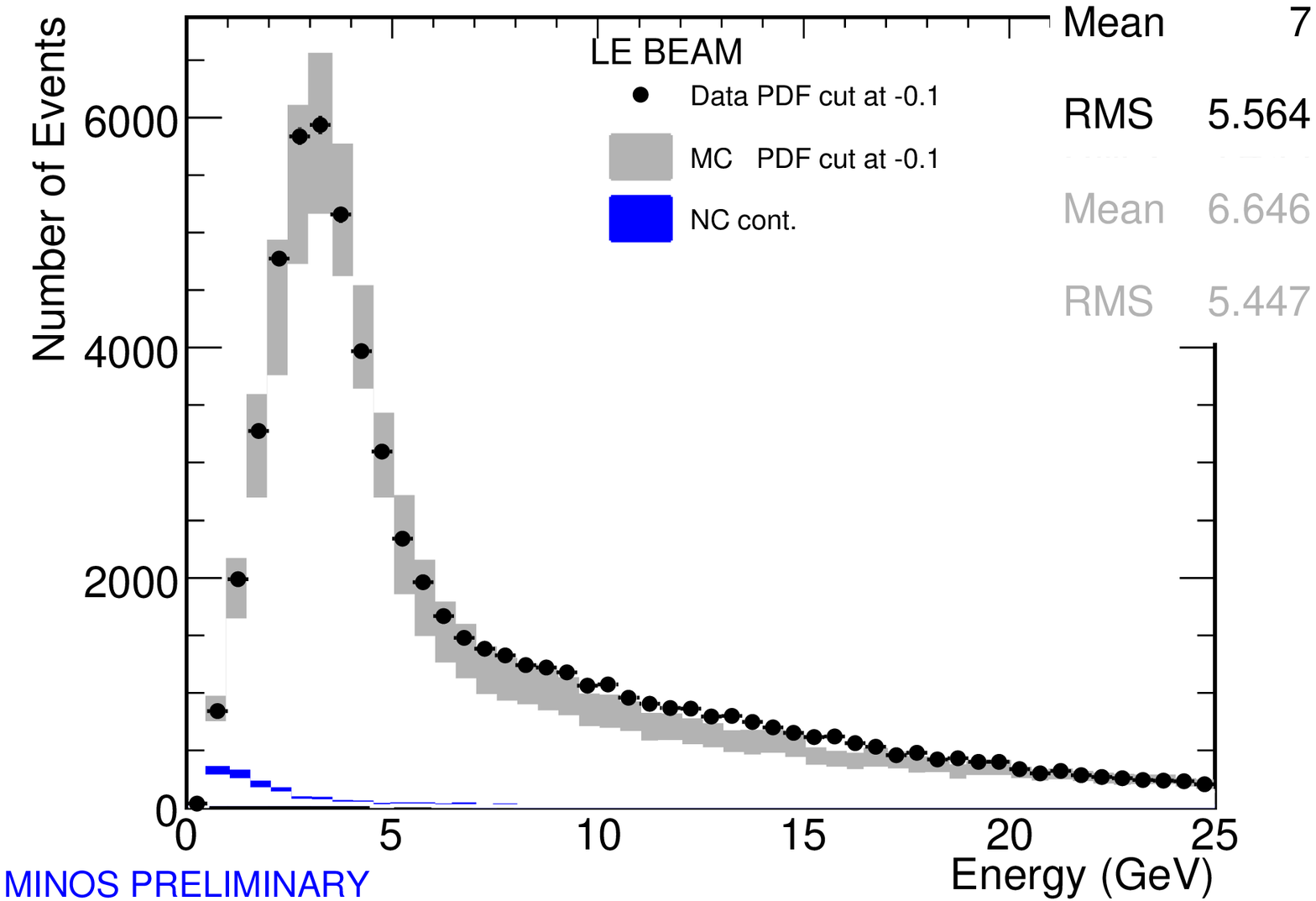}%
  \sizedfig{0.33\textwidth}{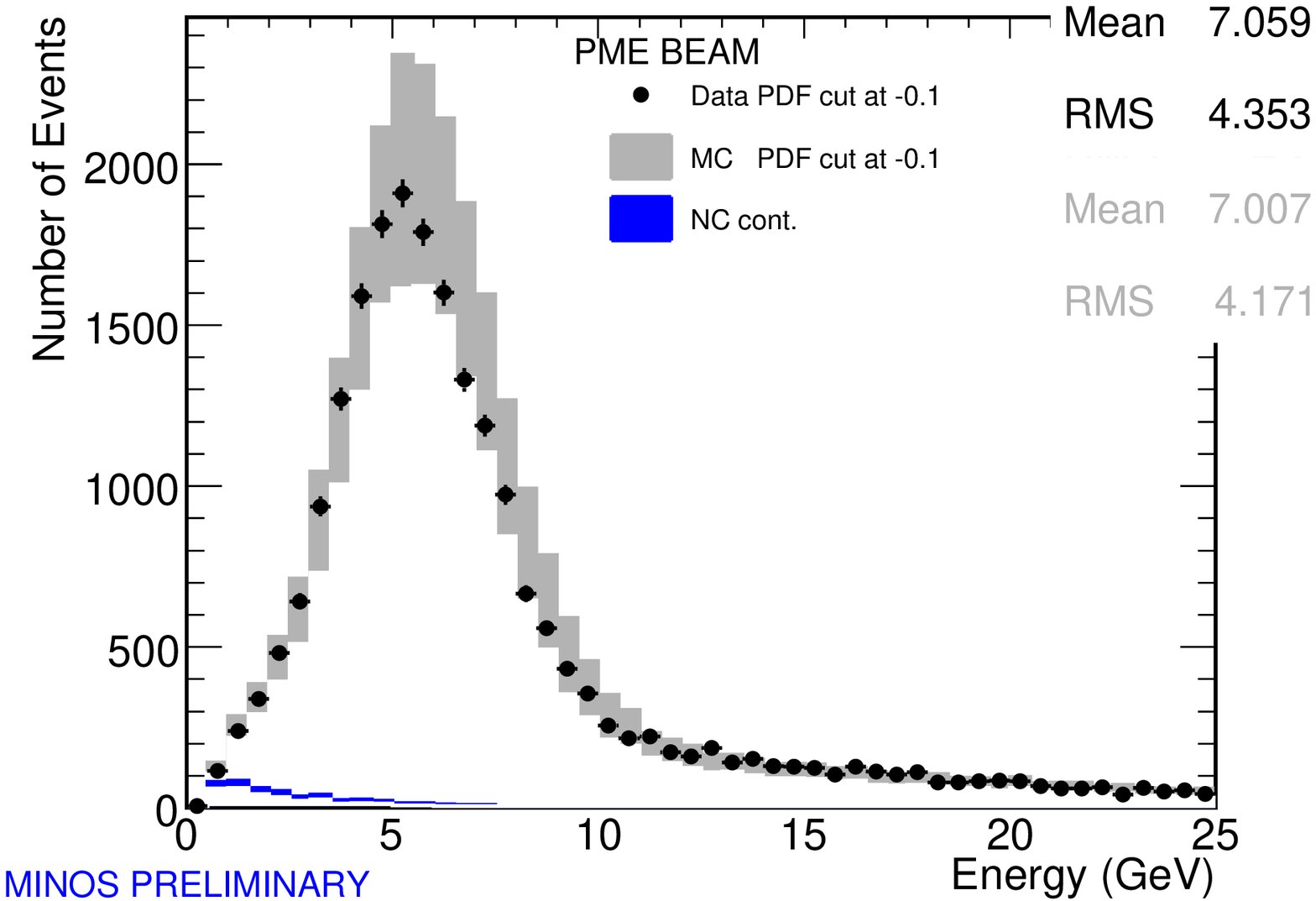}%
  \sizedfig{0.33\textwidth}{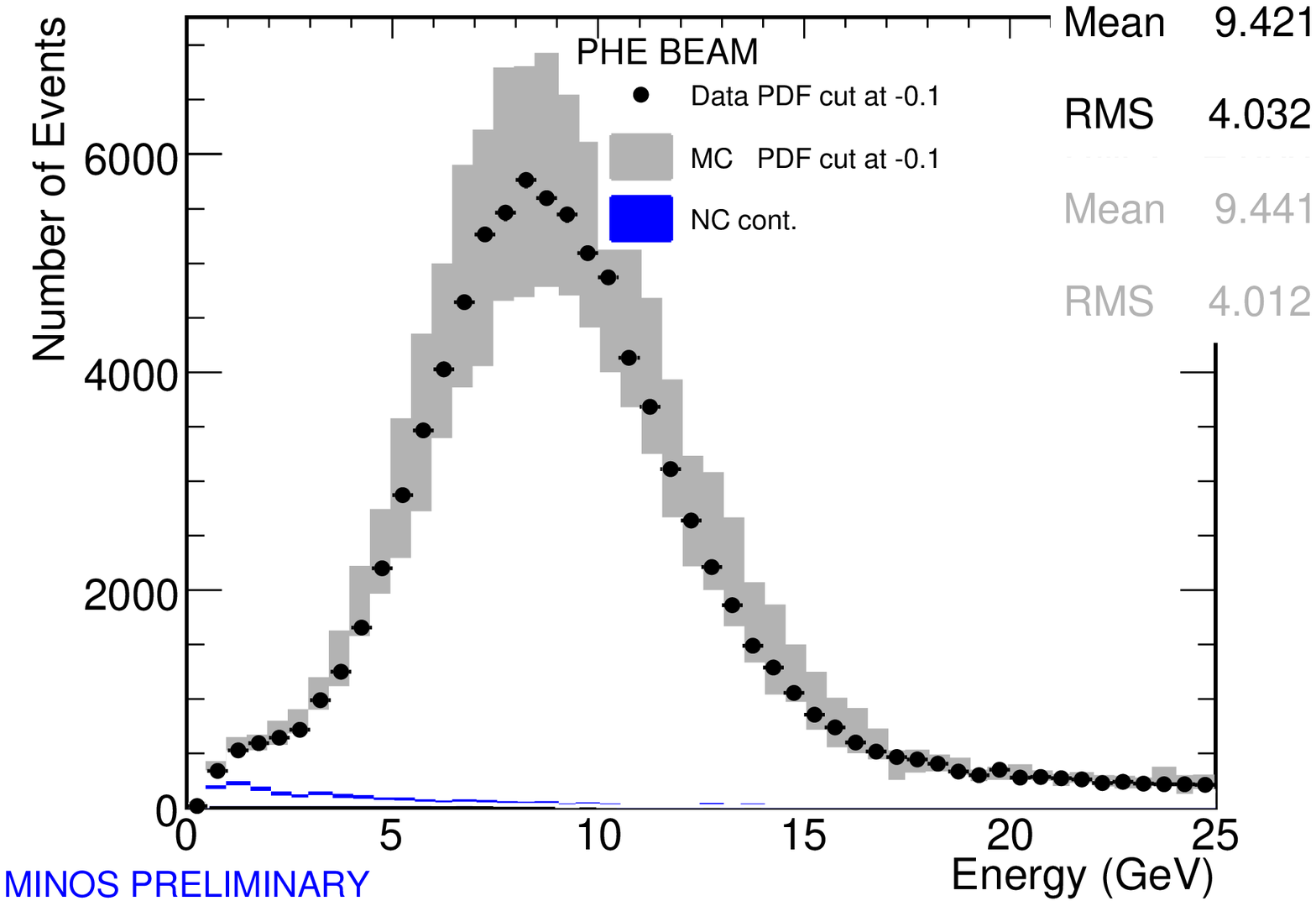}
  \sizedfig{0.33\textwidth}{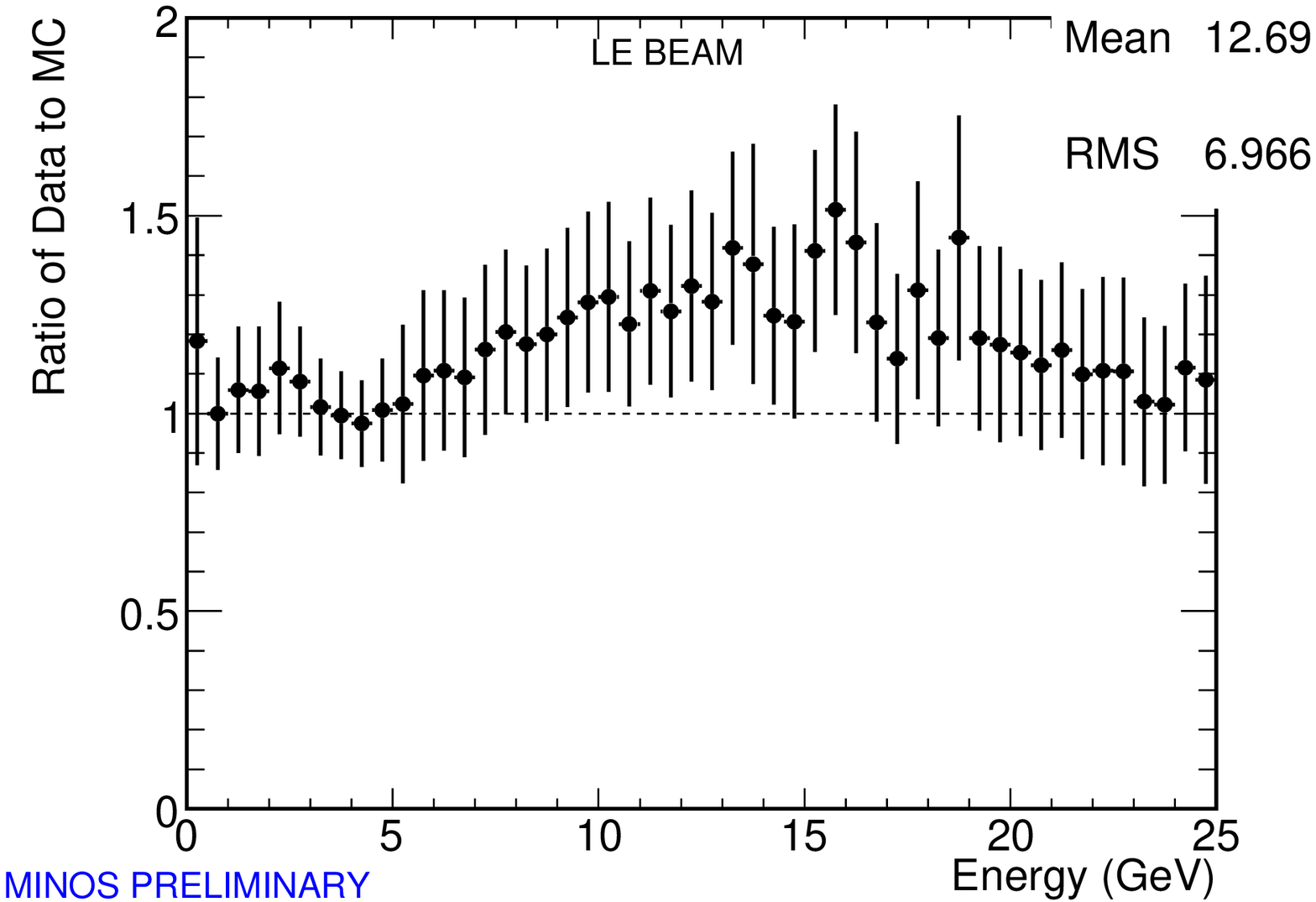}%
  \sizedfig{0.33\textwidth}{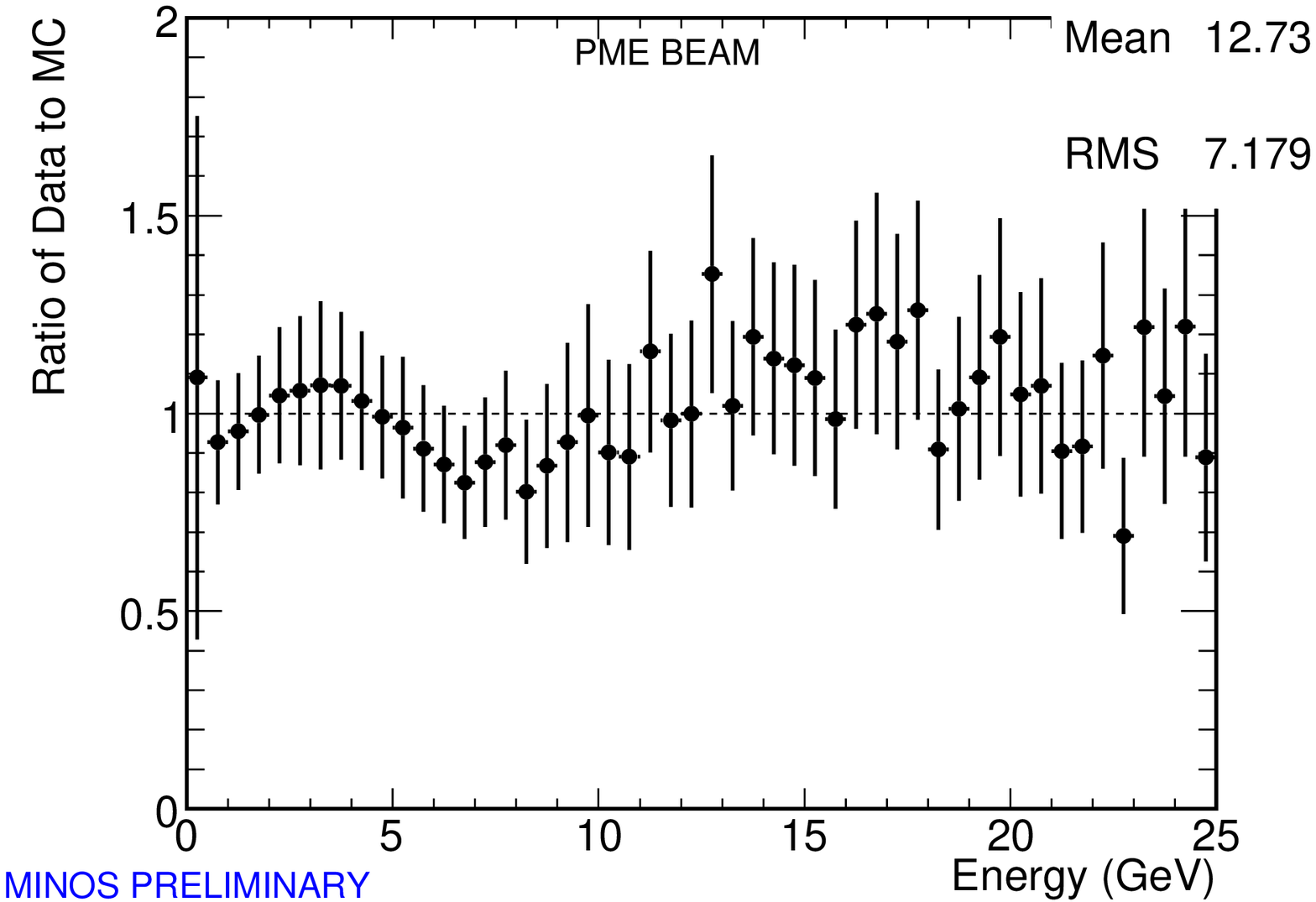}%
  \sizedfig{0.33\textwidth}{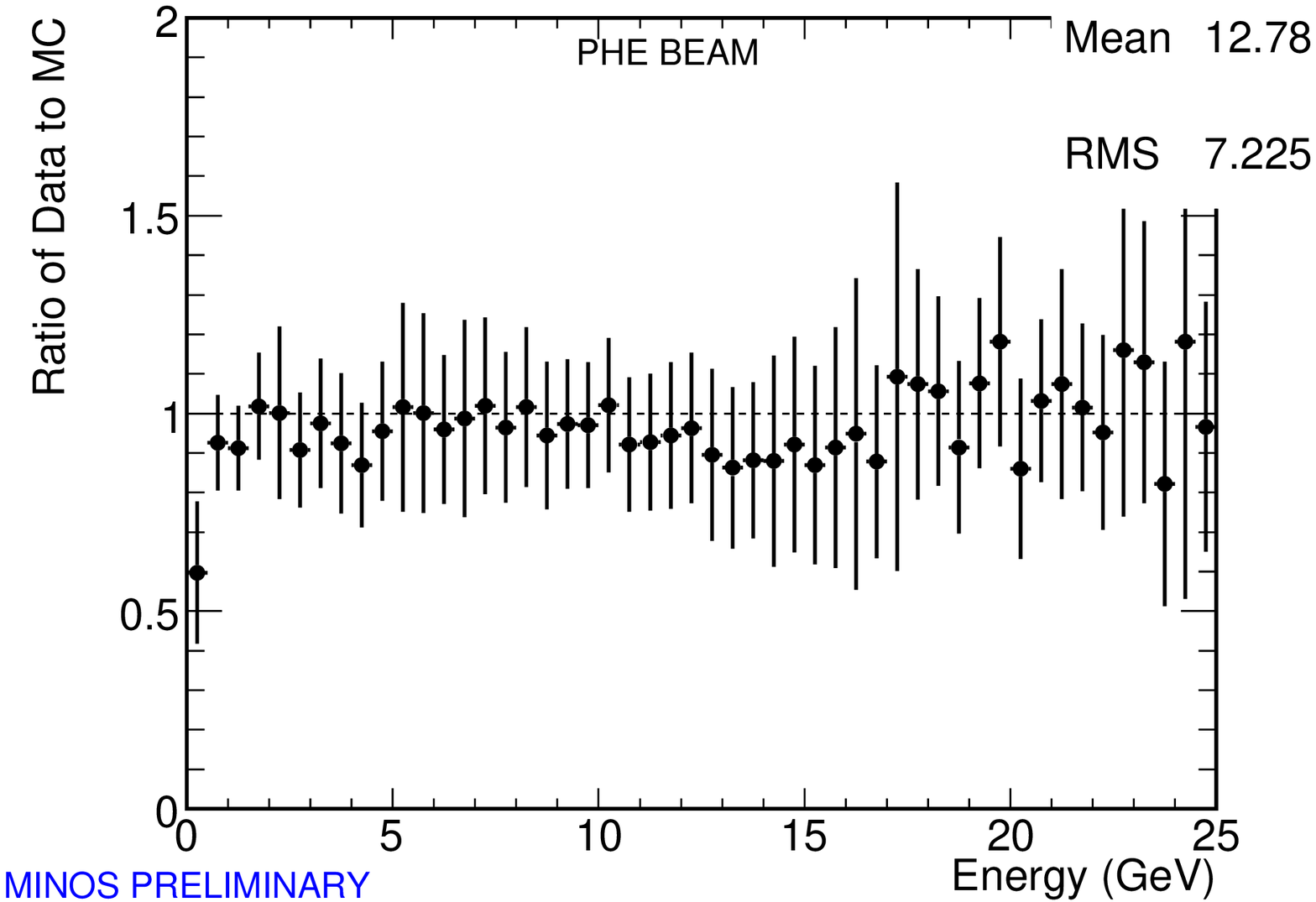}
  \caption{MINOS Near Detector data and GNUMI MC simulation comparison of interaction spectra\cite{wine-and-cheese}.  From left, LE, pME and pHE beams.  Top row shows data spectra as black points, MC as gray shaded region with a width representing MC uncertainty.  Bottom row shows ratio of data to MC.}
  \label{fig:nd-data-compare}
\end{sidewaysfigure}

Figure~\ref{fig:nd-data-compare} shows a comparison between GNUMI MC
simulation and data collected in the MINOS near detector.  There
``LE-10'' corresponds to the ``LE'' configuration in this work.  The
ME simulation used in this work is from a slightly older version (v15)
of GNUMI than is used in this comparison (v18).  No tuning to the
MINOS data has been done and data/MC comparison shows very good
agreement.  

\section{Off-axis Event Rates}

\begin{sidewaysfigure}[htbp]
  \centering
  \sizedfig{0.55\textwidth}{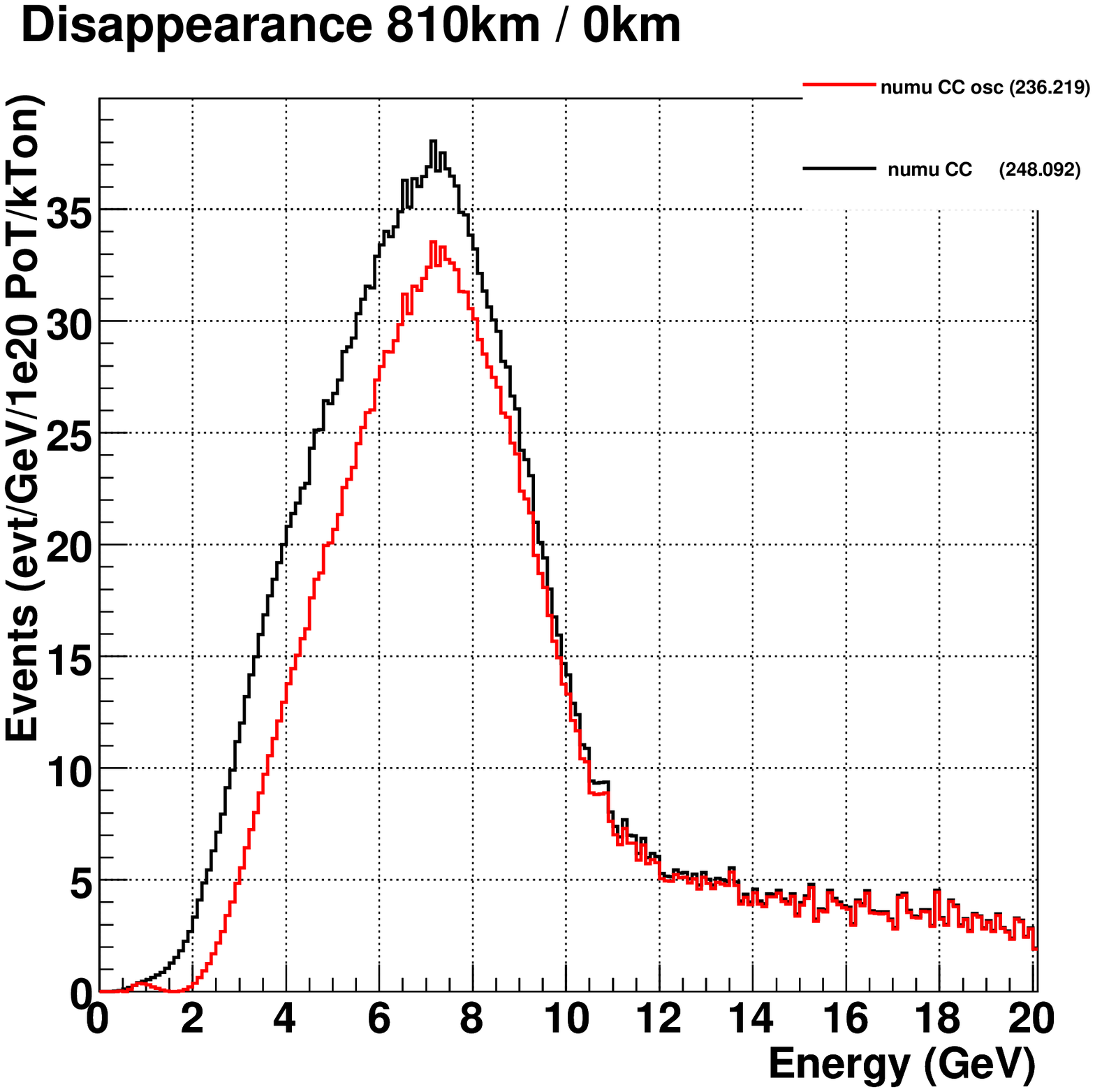}%
  \sizedfig{0.55\textwidth}{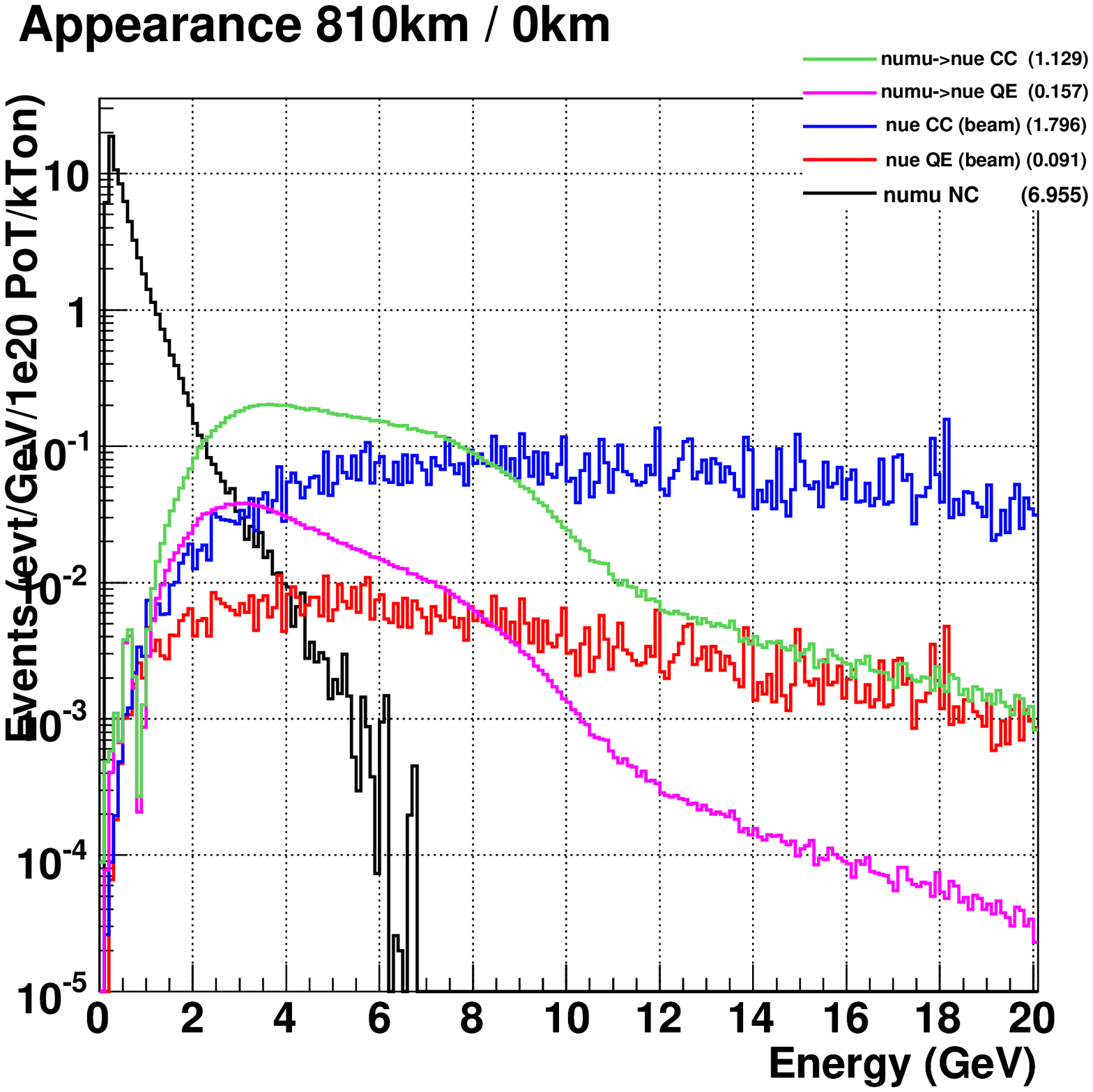}
  \caption{Disappearance (left) and appearance (right) for 810 km baseline and 0 km off-axis.}
  \label{fig:h-exp-810-0}
\end{sidewaysfigure}

\begin{sidewaysfigure}[htbp]
  \centering
  \sizedfig{0.55\textwidth}{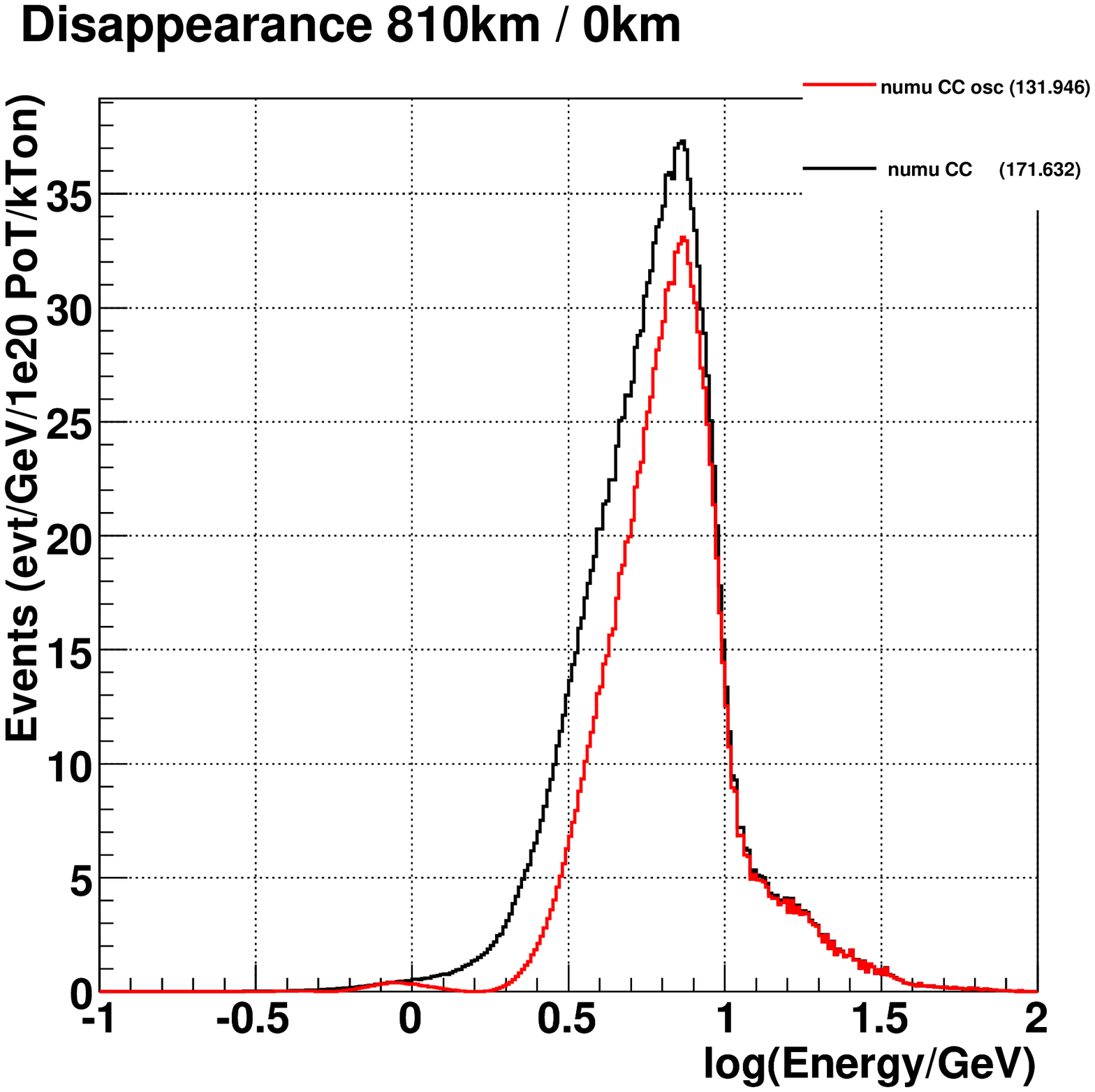}%
  \sizedfig{0.55\textwidth}{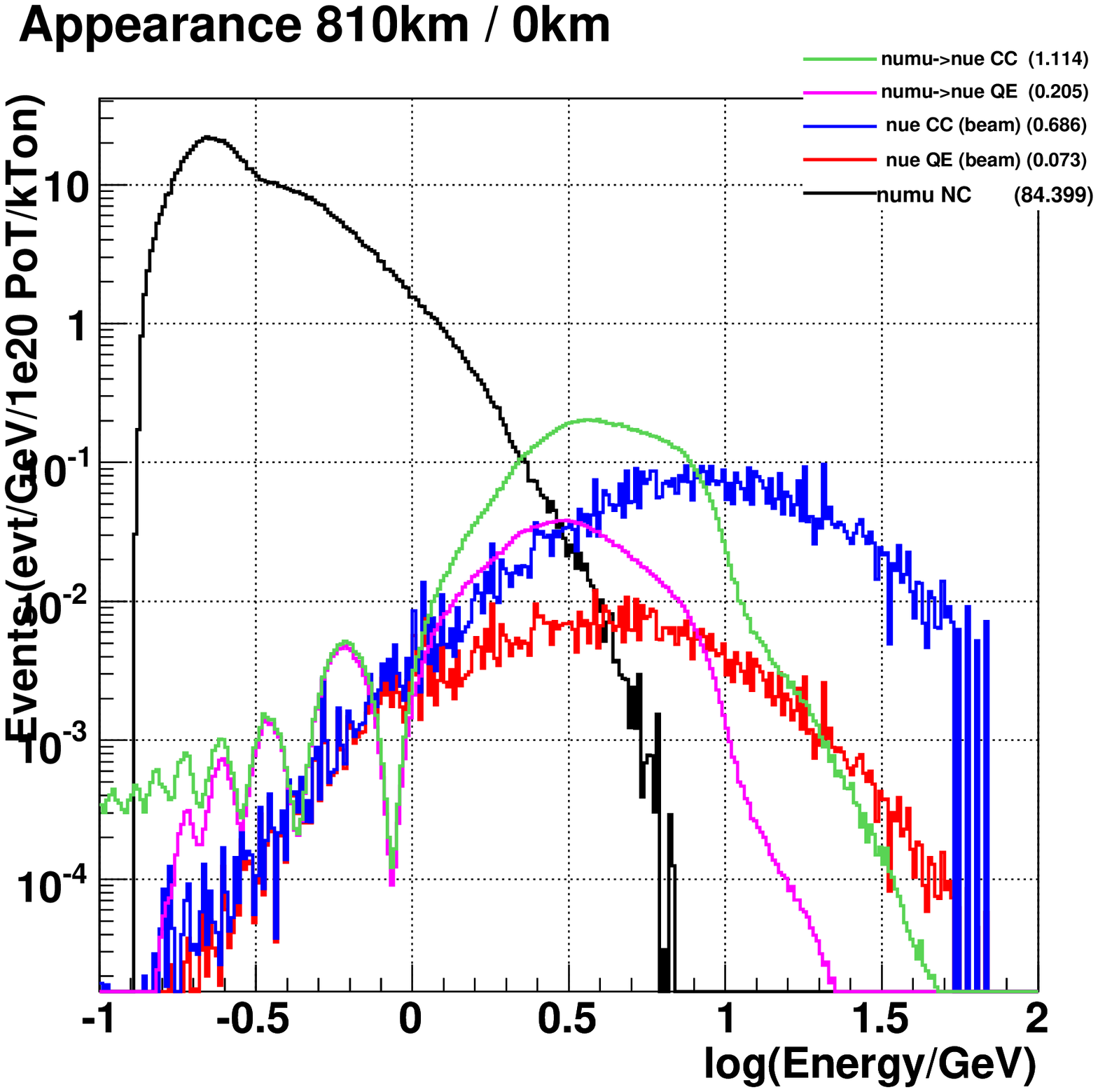}
  \caption{Disappearance (left) and appearance (right) for 810 km baseline and 0 km off-axis.}
  \label{fig:l-exp-810-0}
\end{sidewaysfigure}

\begin{sidewaysfigure}[htbp]
  \centering
  \sizedfig{0.55\textwidth}{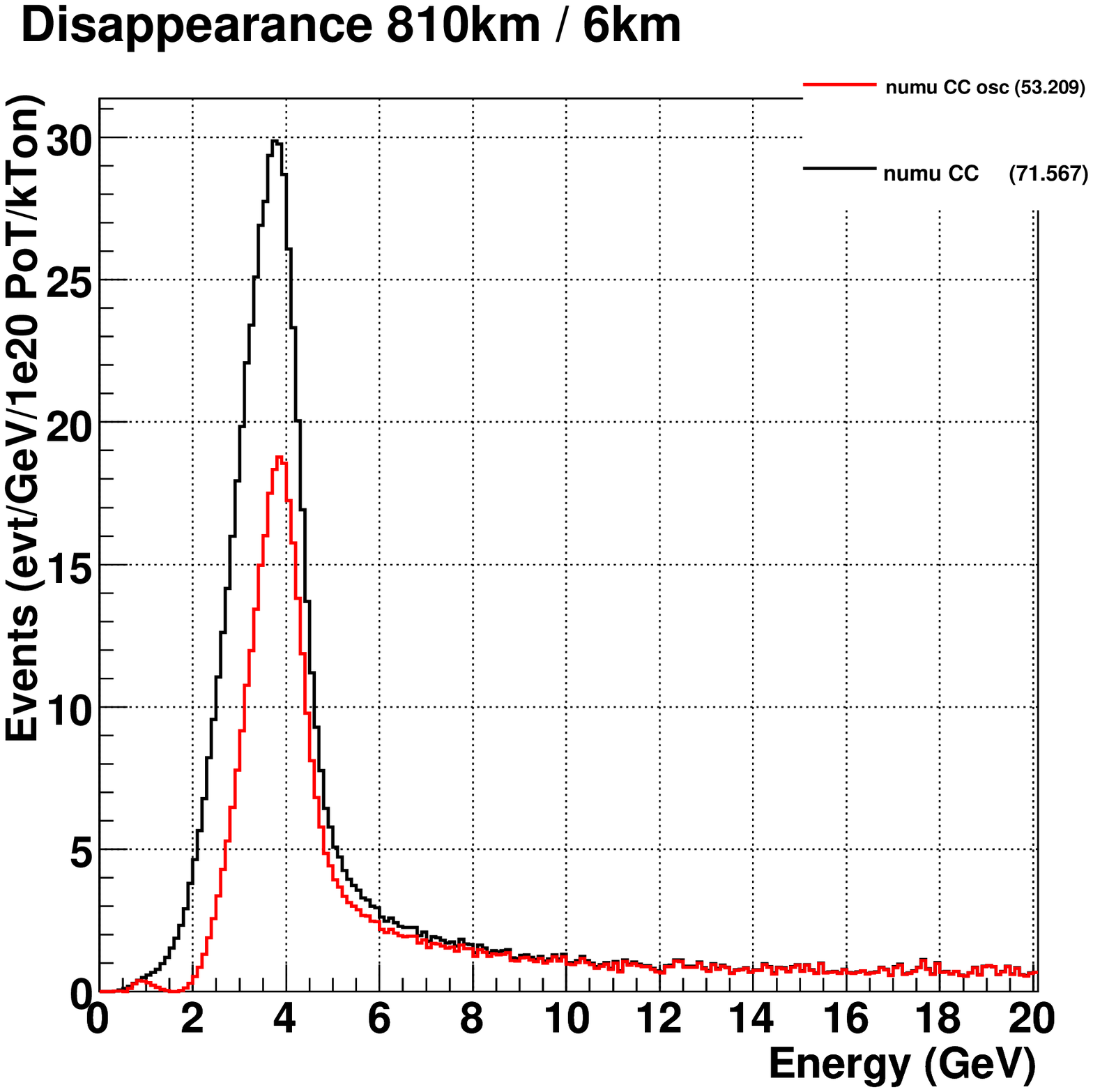}%
  \sizedfig{0.55\textwidth}{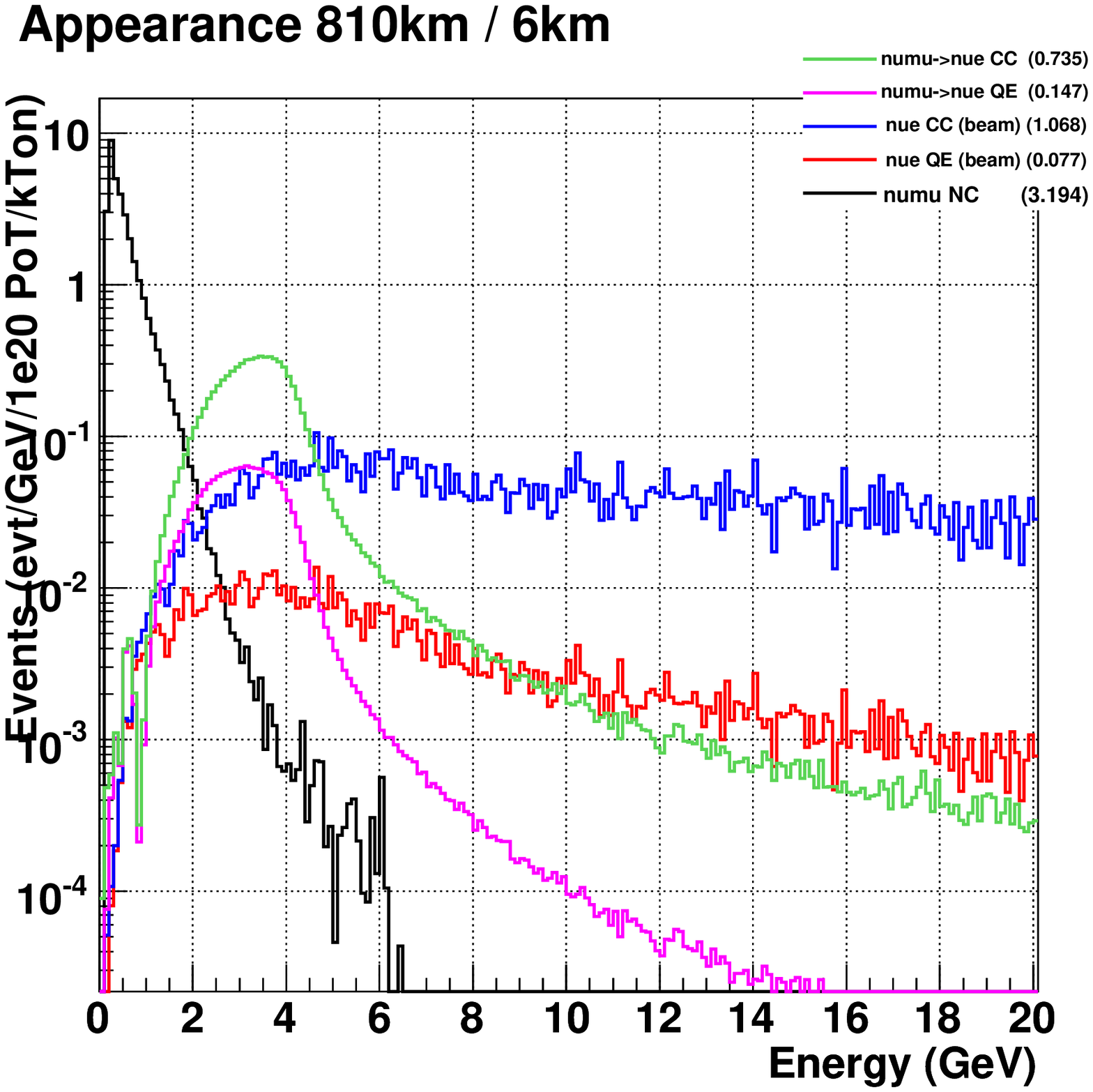}
  \caption{Disappearance (left) and appearance (right) for 810 km baseline and 6 km off-axis.}
  \label{fig:h-exp-810-6}
\end{sidewaysfigure}

\begin{sidewaysfigure}[htbp]
  \centering
  \sizedfig{0.55\textwidth}{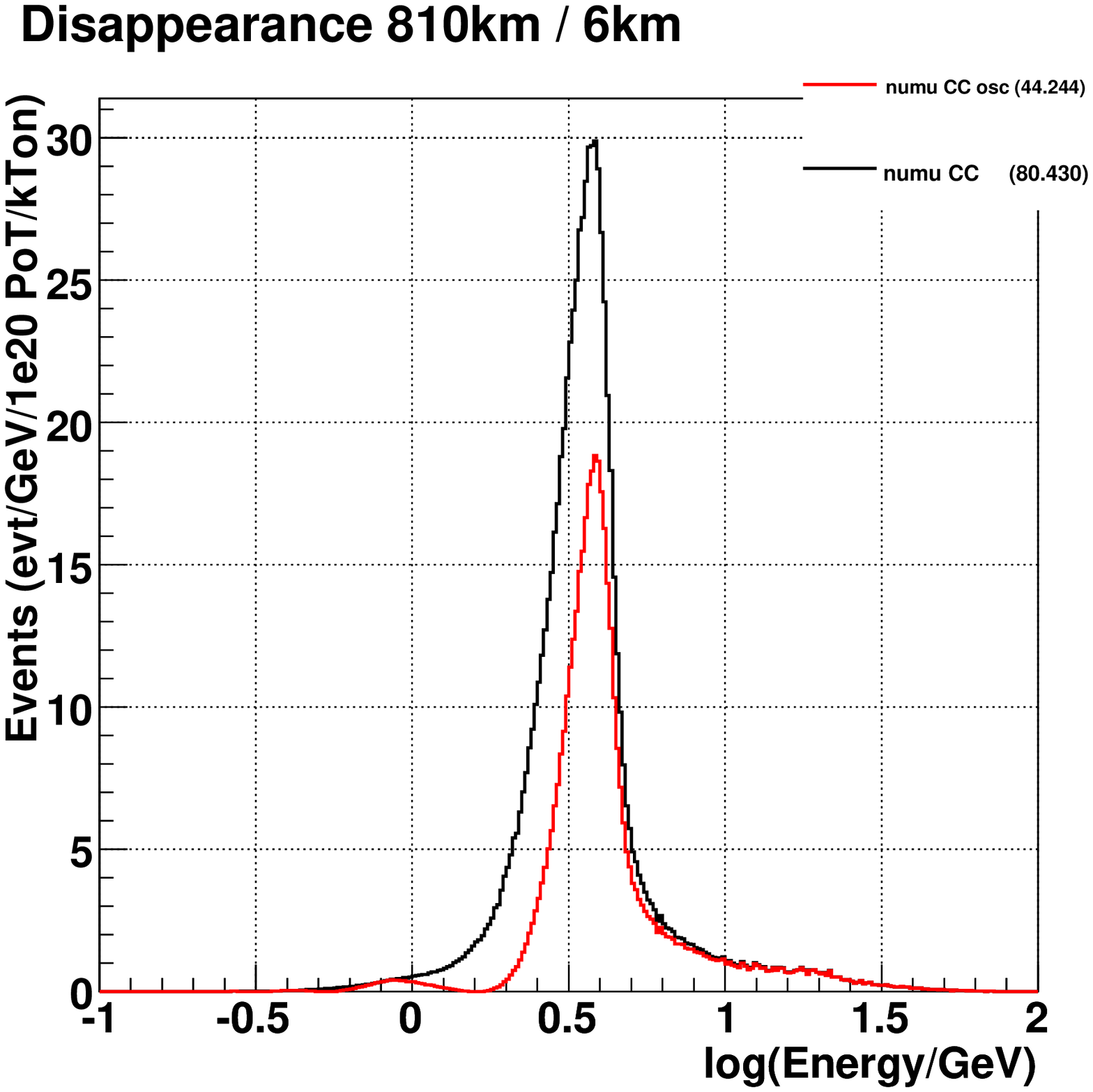}%
  \sizedfig{0.55\textwidth}{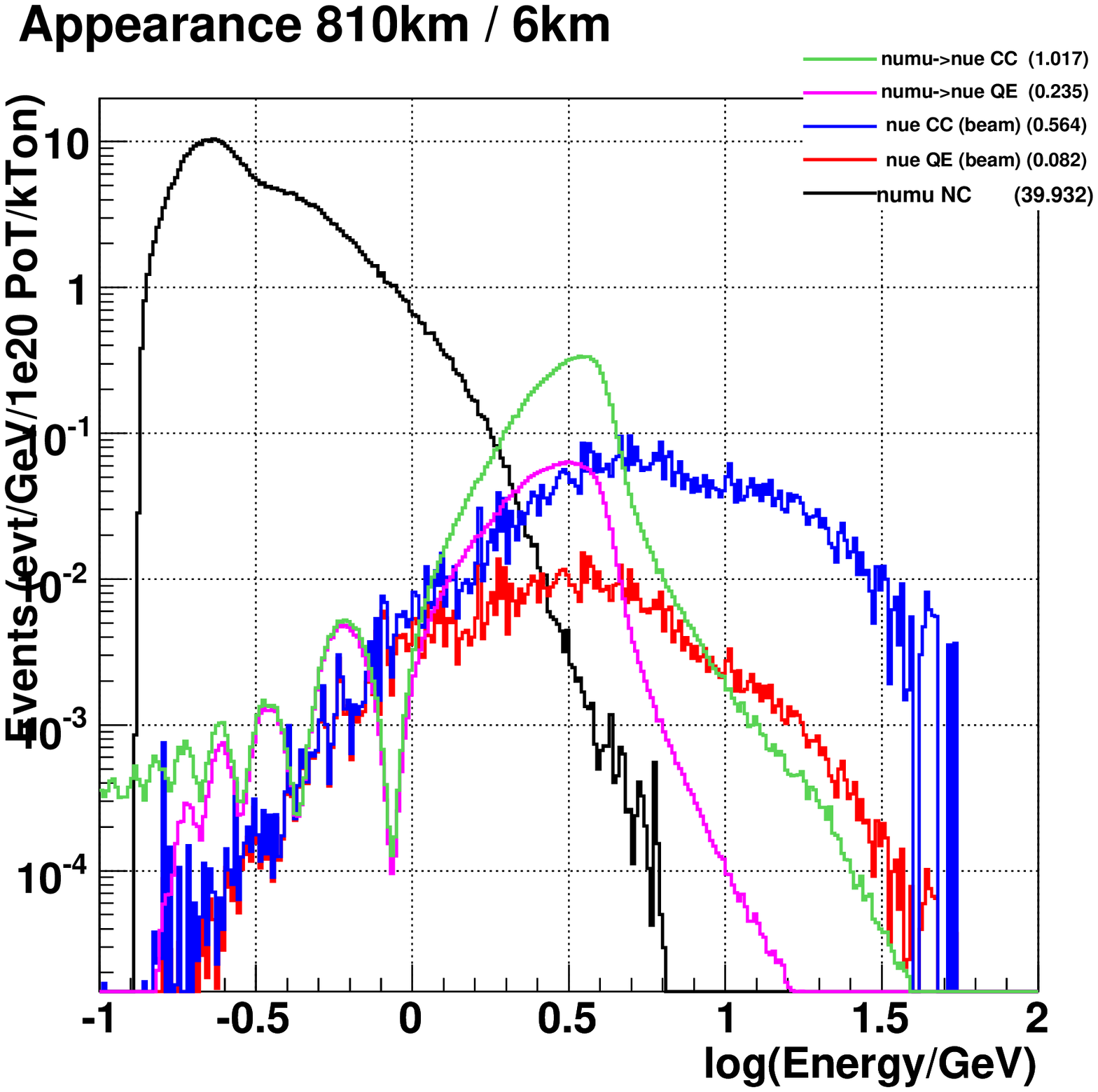}
  \caption{Disappearance (left) and appearance (right) for 810 km baseline and 6 km off-axis.}
  \label{fig:l-exp-810-6}
\end{sidewaysfigure}

\begin{sidewaysfigure}[htbp]
  \centering
  \sizedfig{0.55\textwidth}{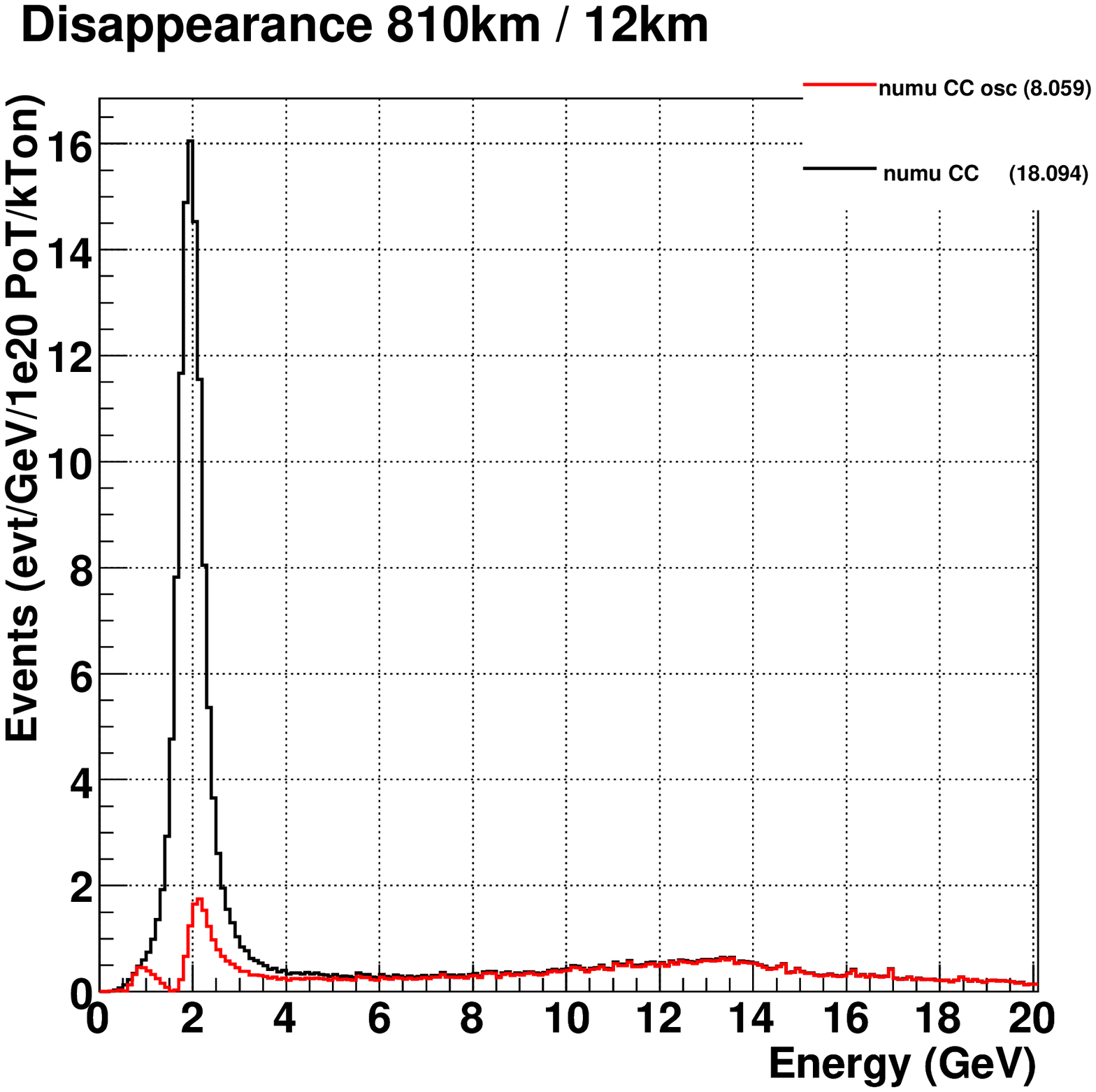}%
  \sizedfig{0.55\textwidth}{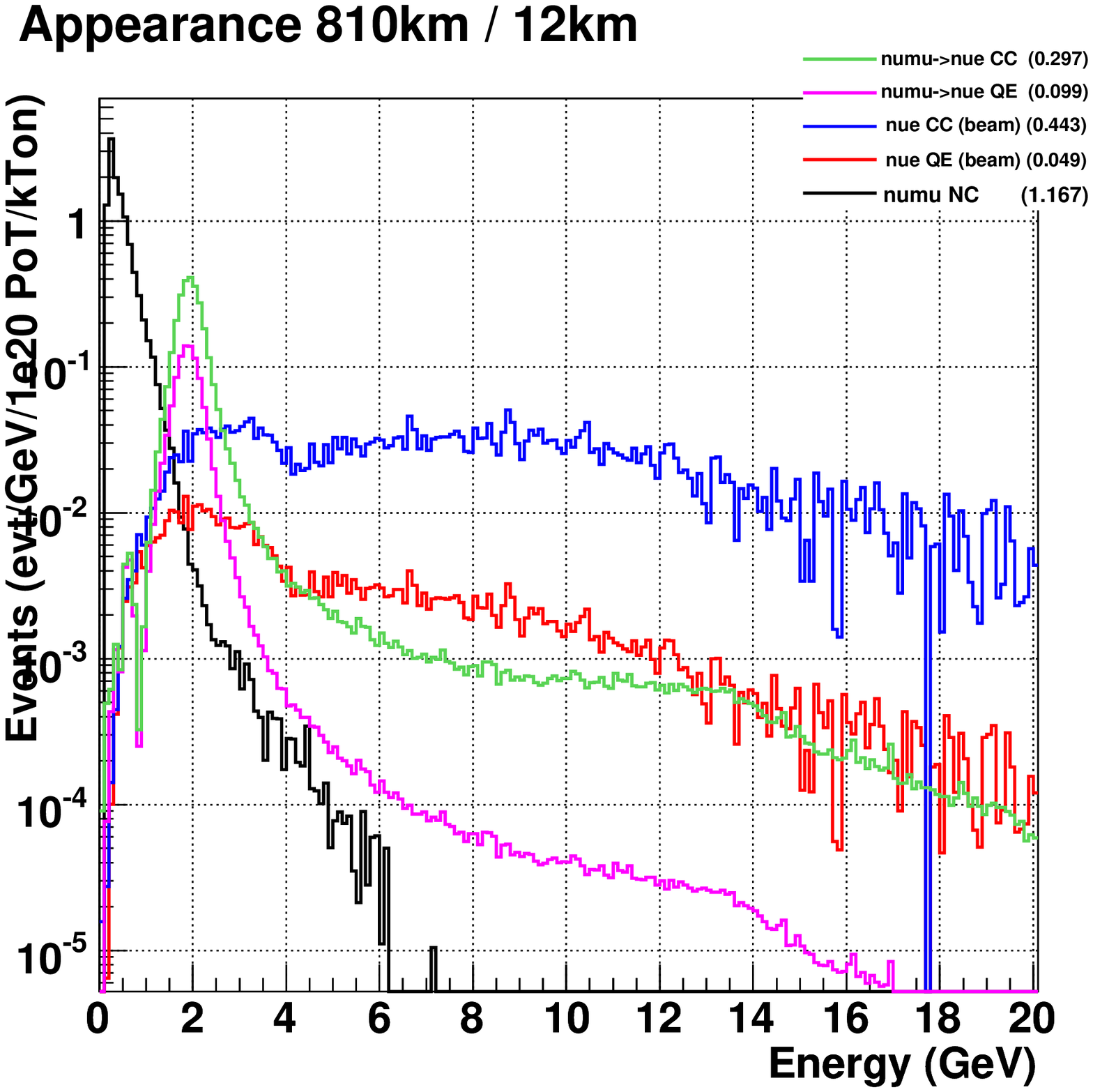}
  \caption{Disappearance (left) and appearance (right) for 810 km baseline and 12 km off-axis.}
  \label{fig:h-exp-810-12}
\end{sidewaysfigure}

\begin{sidewaysfigure}[htbp]
  \centering
  \sizedfig{0.55\textwidth}{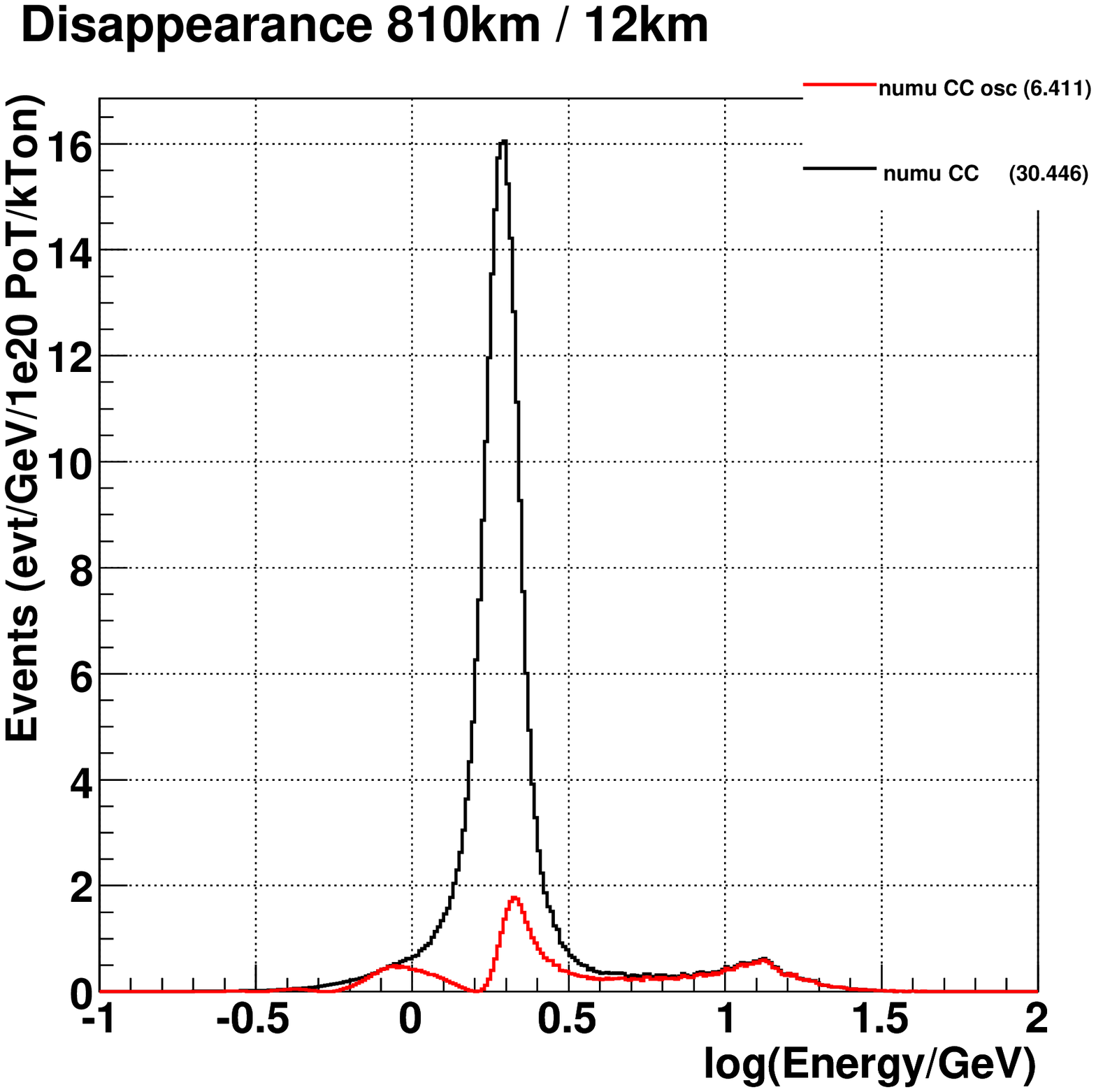}%
  \sizedfig{0.55\textwidth}{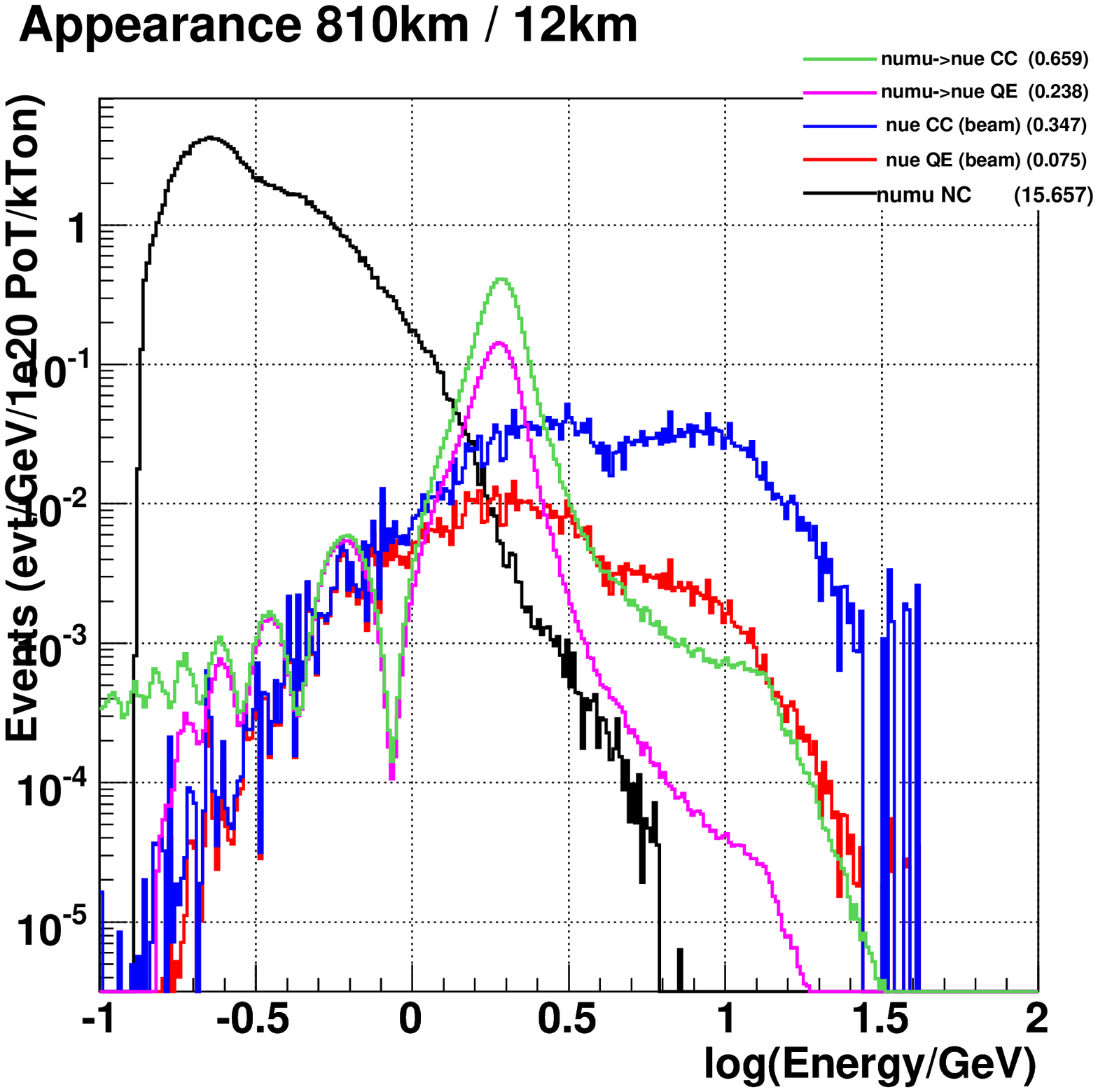}
  \caption{Disappearance (left) and appearance (right) for 810 km baseline and 12 km off-axis.}
  \label{fig:l-exp-810-12}
\end{sidewaysfigure}

\begin{sidewaysfigure}[htbp]
  \centering
  \sizedfig{0.55\textwidth}{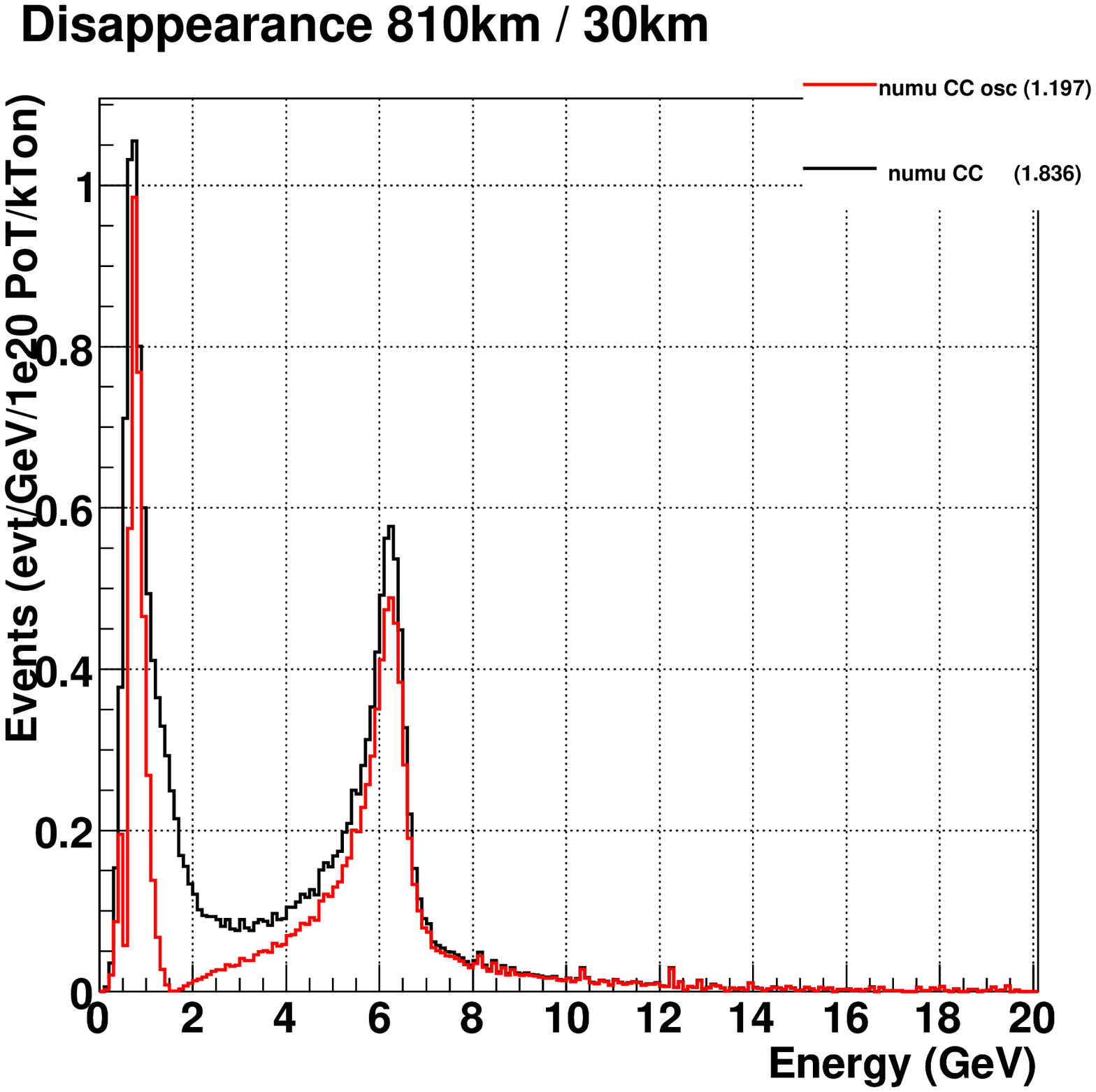}%
  \sizedfig{0.55\textwidth}{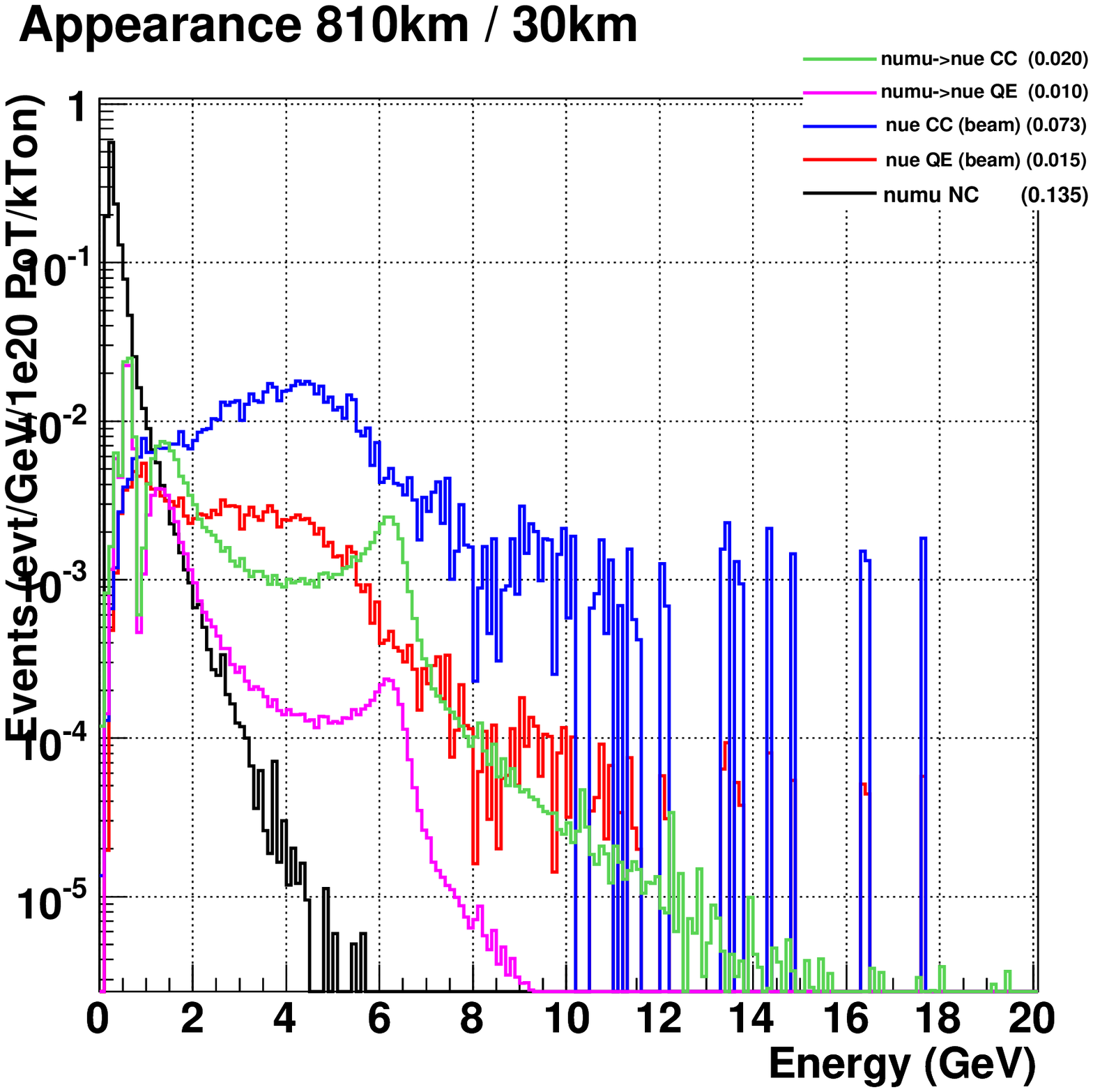}
  \caption{Disappearance (left) and appearance (right) for 810 km baseline and 30 km off-axis.}
  \label{fig:h-exp-810-30}
\end{sidewaysfigure}

\begin{sidewaysfigure}[htbp]
  \centering
  \sizedfig{0.55\textwidth}{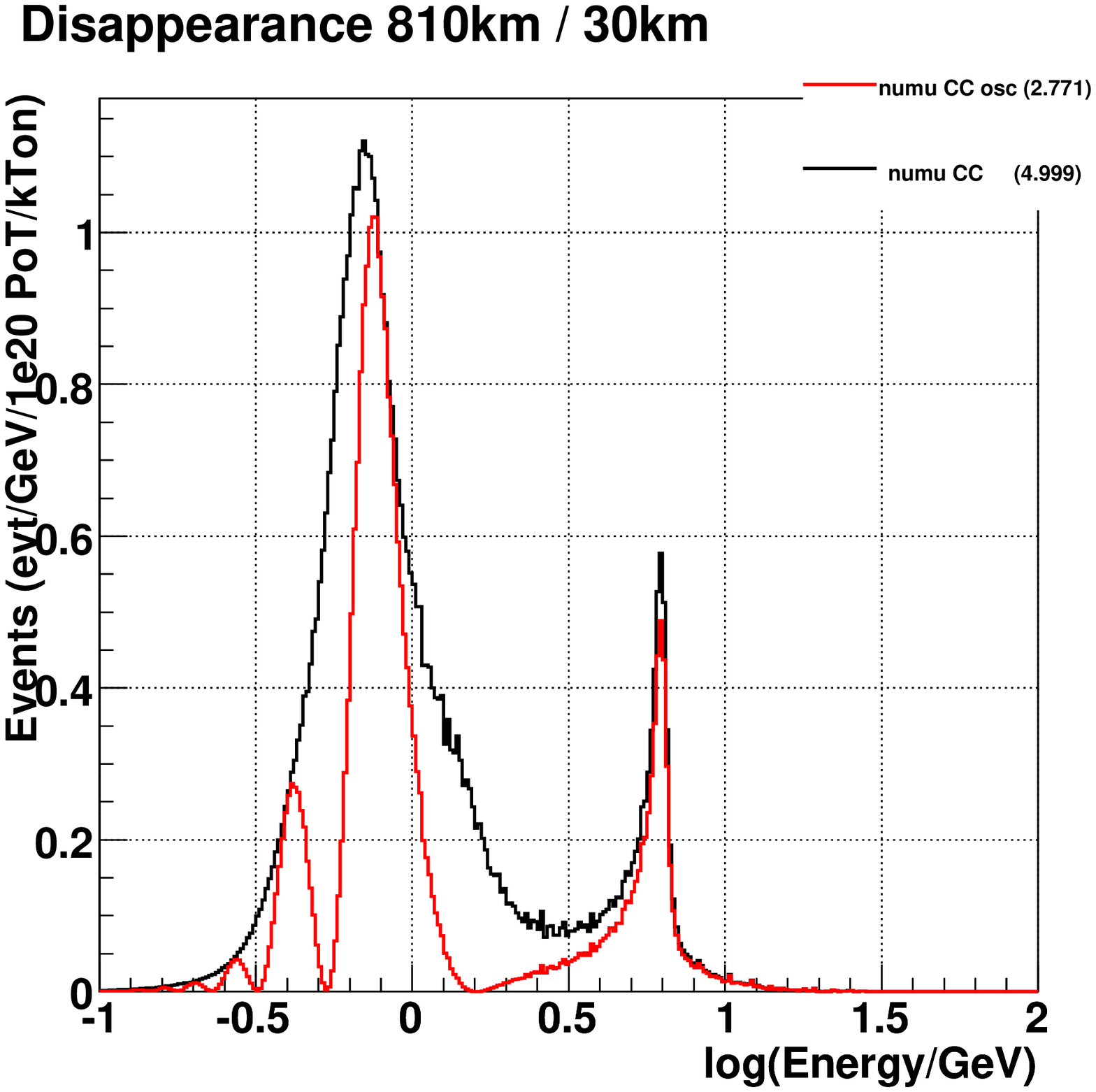}%
  \sizedfig{0.55\textwidth}{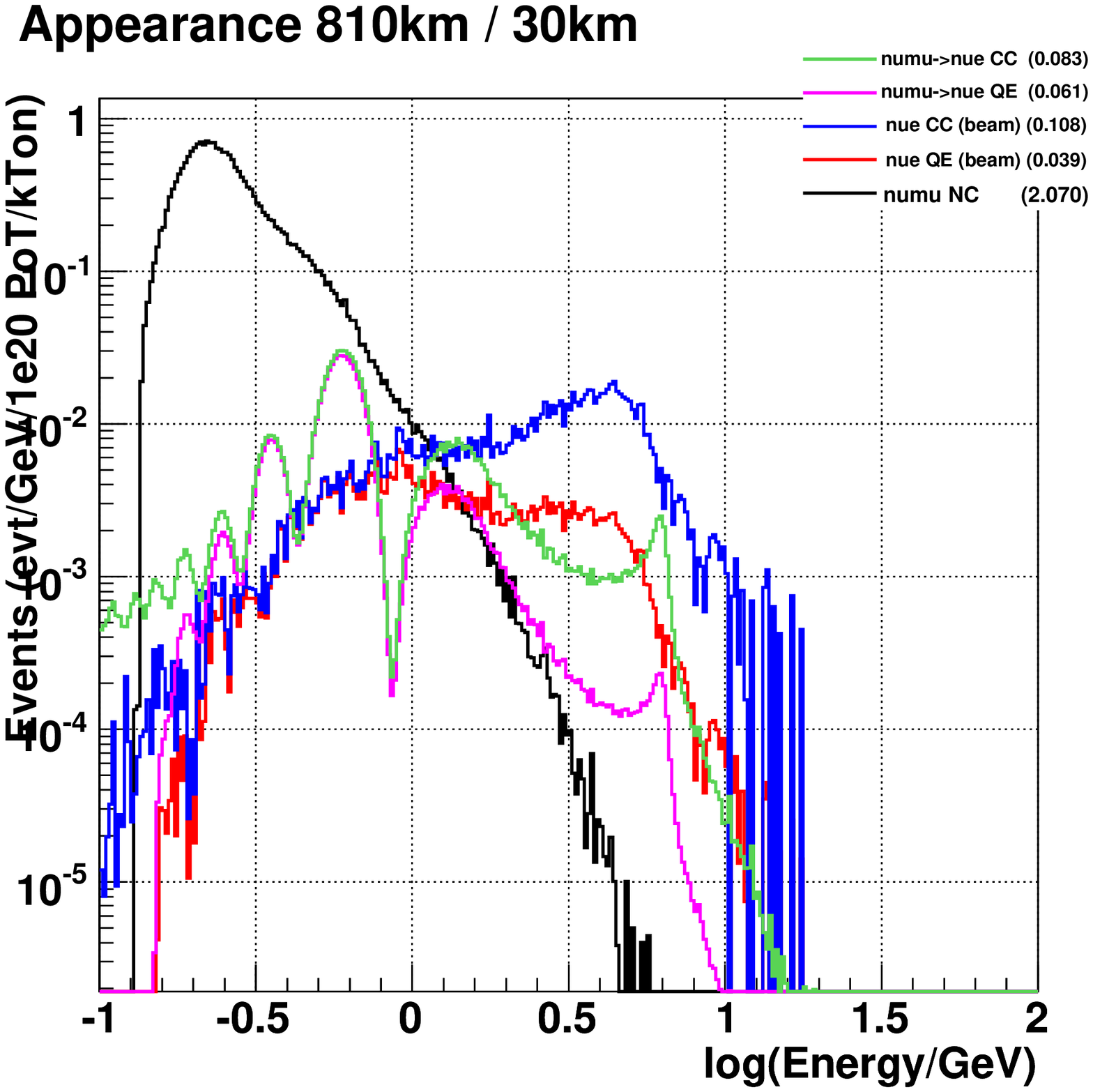}
  \caption{Disappearance (left) and appearance (right) for 810 km baseline and 30 km off-axis.}
  \label{fig:l-exp-810-30}
\end{sidewaysfigure}

\begin{sidewaysfigure}[htbp]
  \centering
  \sizedfig{0.55\textwidth}{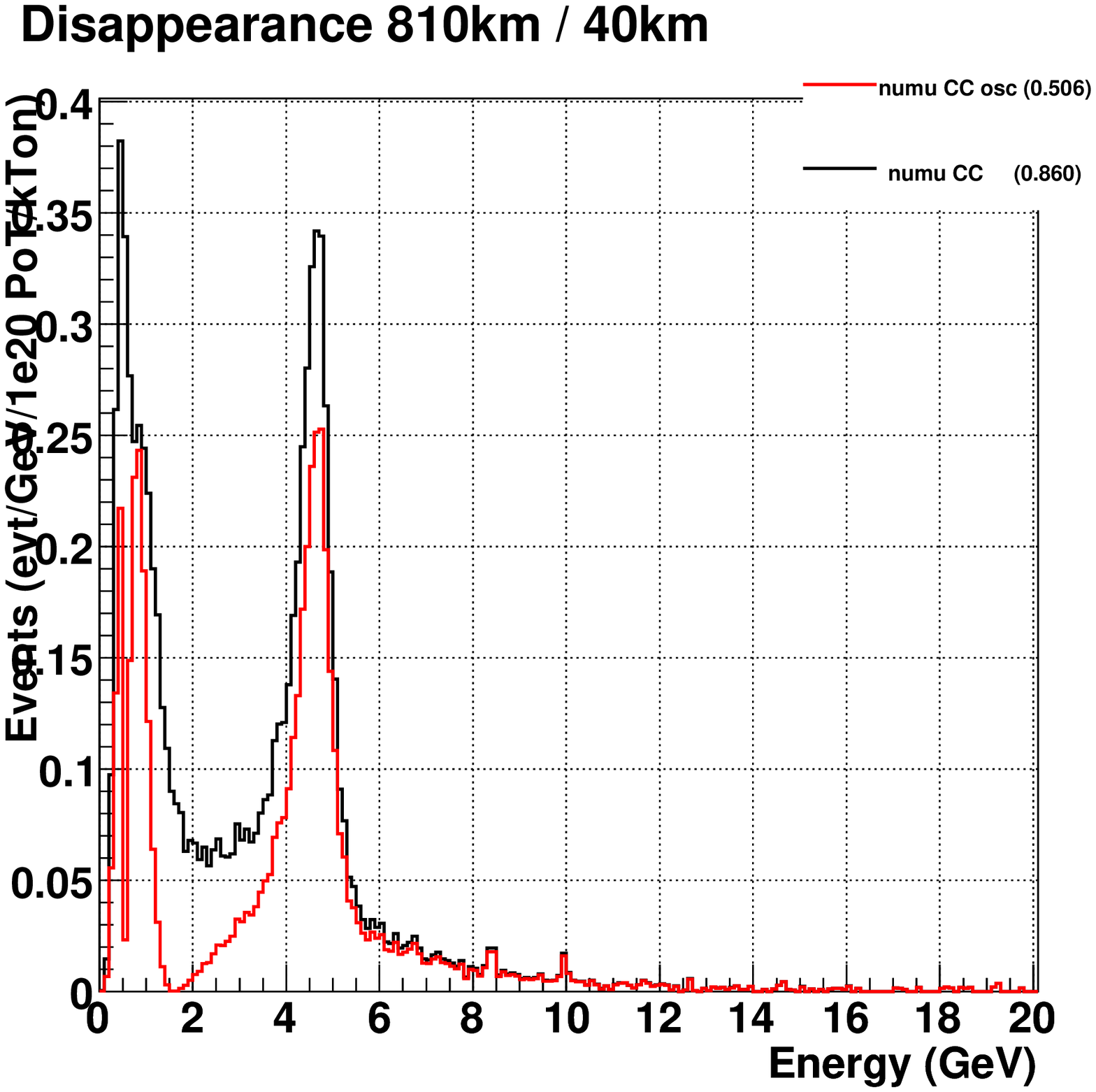}%
  \sizedfig{0.55\textwidth}{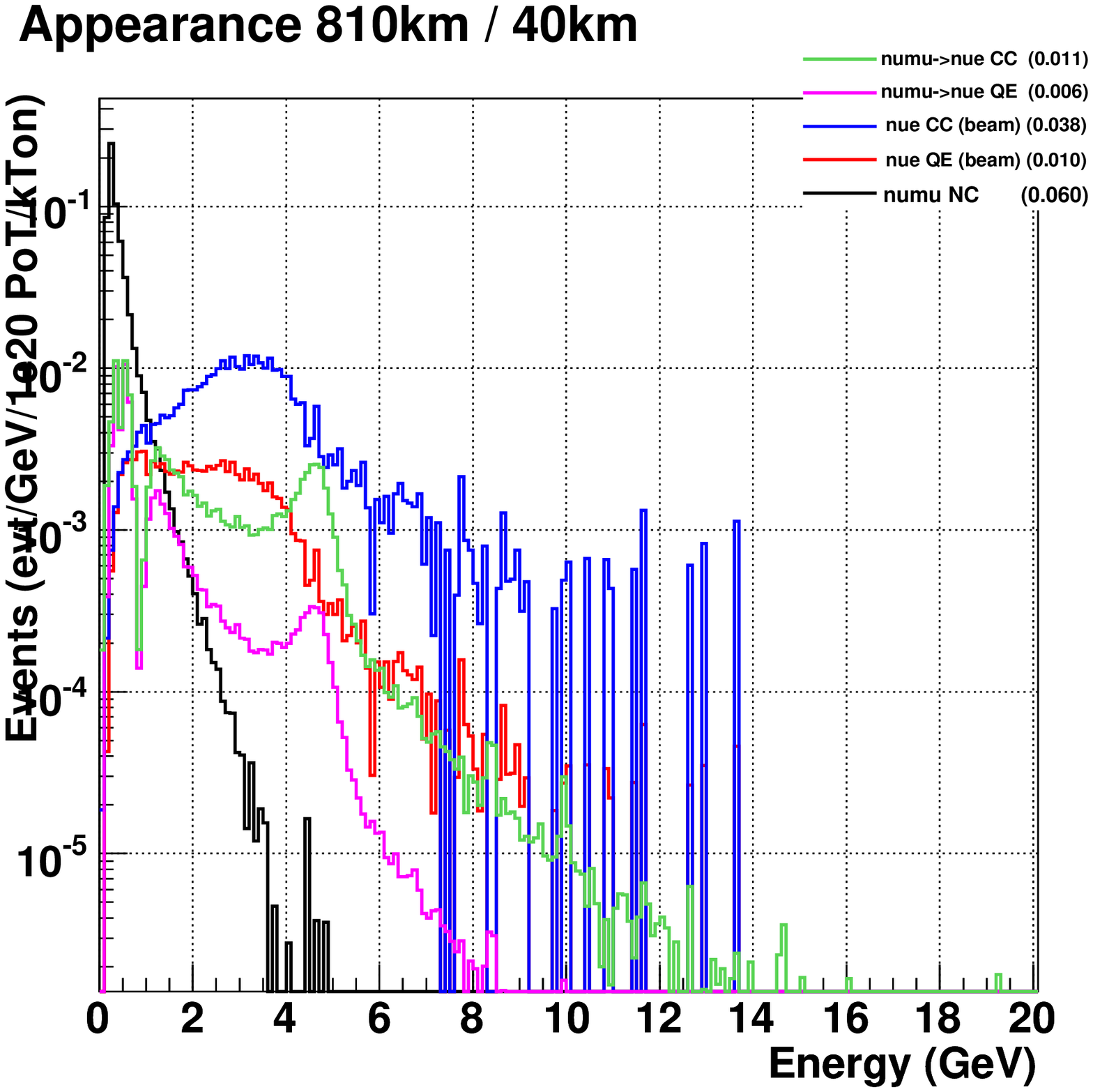}
  \caption{Disappearance (left) and appearance (right) for 810 km baseline and 40 km off-axis.}
  \label{fig:h-exp-810-40}
\end{sidewaysfigure}

\begin{sidewaysfigure}[htbp]
  \centering
  \sizedfig{0.55\textwidth}{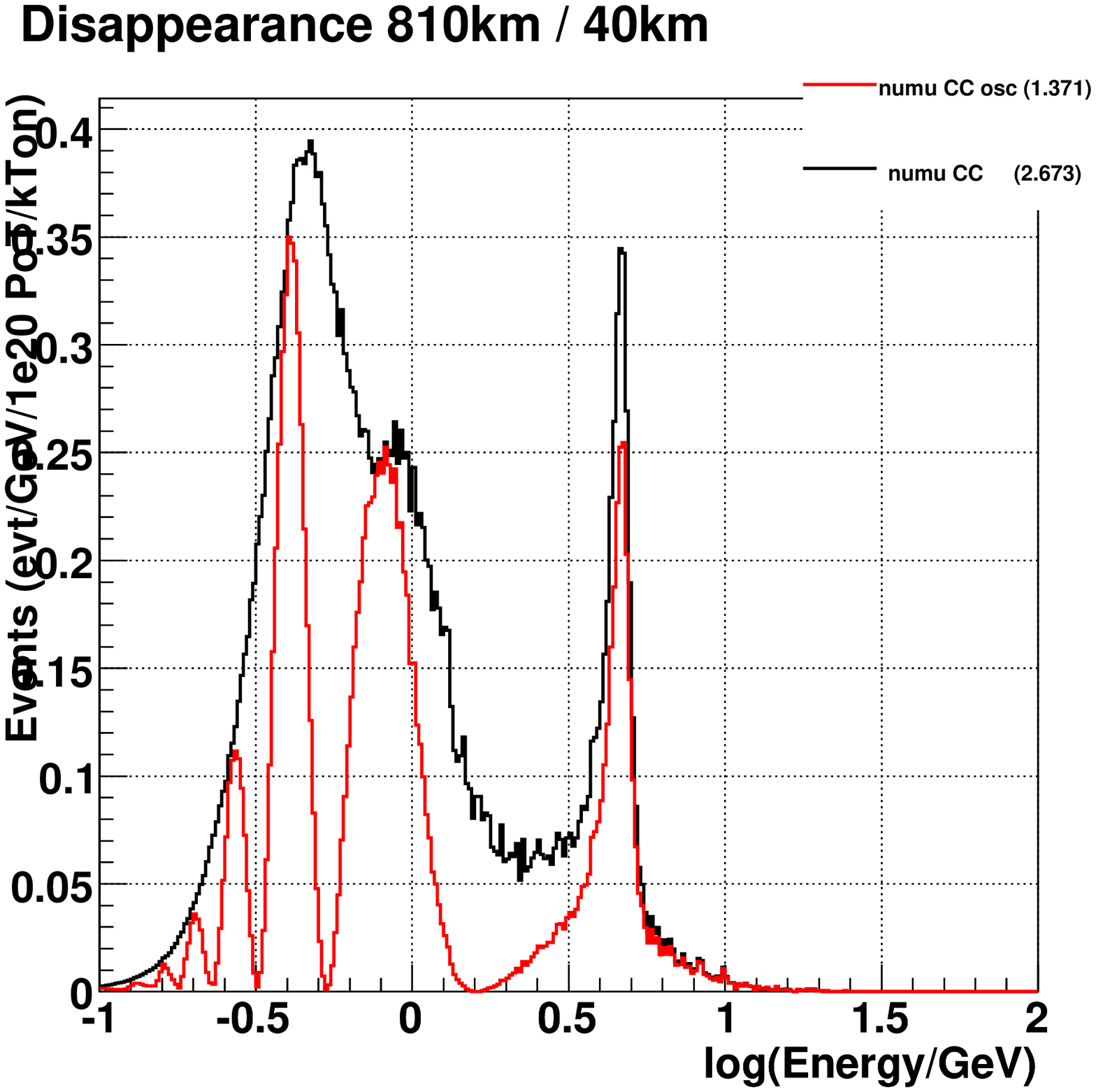}%
  \sizedfig{0.55\textwidth}{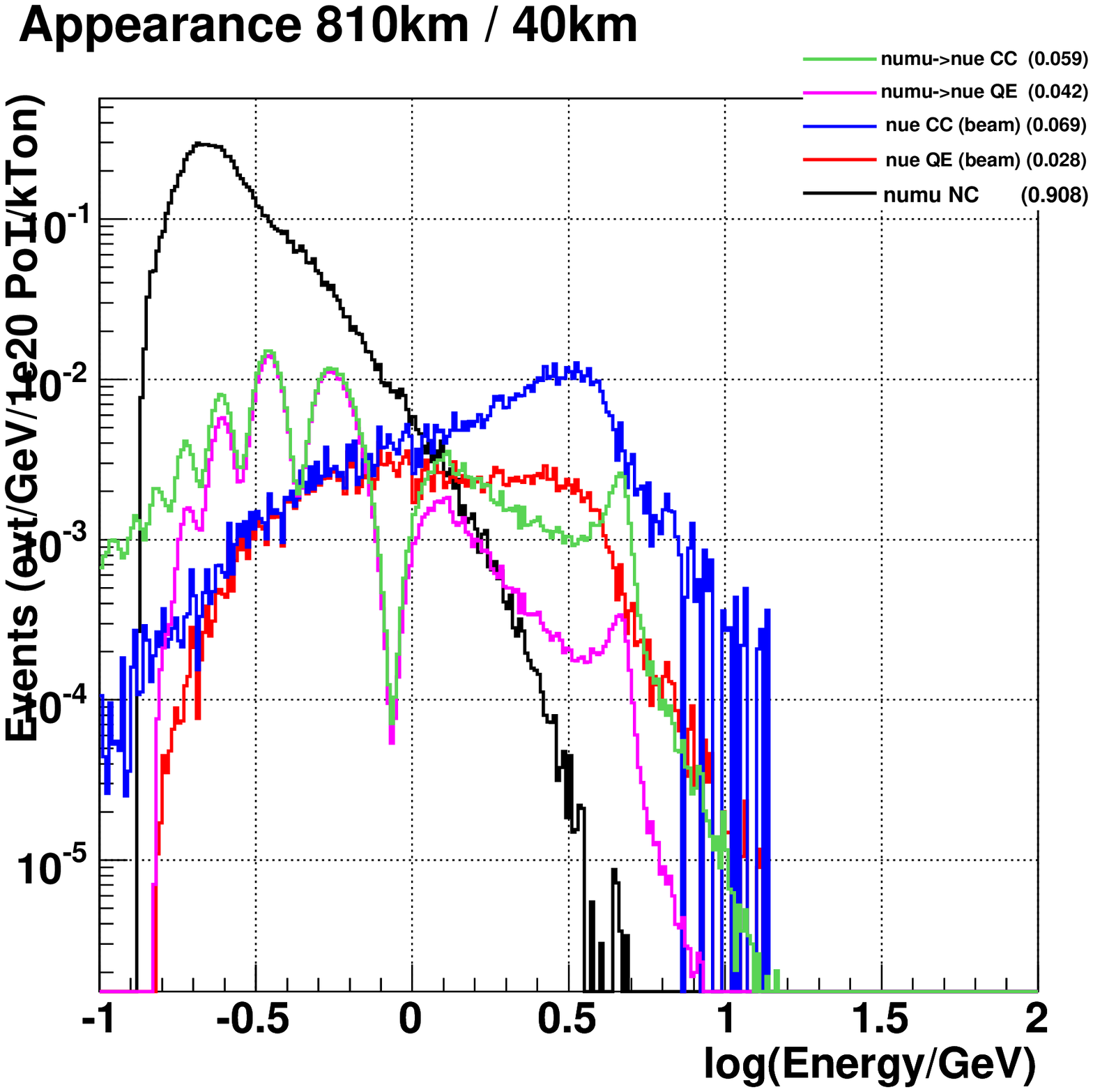}
  \caption{Disappearance (left) and appearance (right) for 810 km baseline and 40 km off-axis.}
  \label{fig:l-exp-810-40}
\end{sidewaysfigure}

All of the above is combined to estimate interaction spectra and rates
for disappearance and appearance modes for detectors placed in the
NuMI beam.  Figures~\ref{fig:h-exp-810-0}-\ref{fig:l-exp-810-40} show
these spectra for an 810km baseline at 0, 6, 12, 30 and 40 km off
axis.  The event rates are summaries in Table~\ref{tab:rates}.

\begin{table}[htbp]
  \centering
  \begin{tabular}{|r|r@{.}l|r@{.}l|r@{.}l|r@{.}l|r@{.}l|r@{.}l|r@{.}l|}
    \hline
    km o.a. & \multicolumn{2}{c|}{\numu{} CC} & \multicolumn{2}{c|}{\numu{} CC osc} & \multicolumn{2}{c|}{\nue{} CC beam} & \multicolumn{2}{c|}{\nue{} QE beam} & \multicolumn{2}{c|}{NC-$1\pi^0$} & \multicolumn{2}{c|}{\numunue{} CC} & \multicolumn{2}{c|}{\numunue{} QE} \\
    \hline
    0       & 248&0      & 225&0          & 1&80           & 0&0914         & 6&96        & 1&40          & 0&188 \\
    6       & 71&6       & 47&0           & 1&068          & 0&0770         & 3&194       & 0&879         & 0&171 \\
    12      & 18&1       & 7&33           & 0&443          & 0&0485         & 1&168       & 0&305         & 0&099 \\
    30      & 1&84       & 1&12           & 0&0730         & 0&0152         & 0&135       & 0&0216        & 0&0108 \\
    40      & 0&860      & 0&479          & 0&0378         & 0&0097         & 0&0605      & 0&0121        & 0&0057 \\
    \hline
  \end{tabular}
  \caption{Summary of event rates per-kTon per-$10^{20}$ PoT for detectors placed 810 km from the target and various distances off-axis (o.a.).}
  \label{tab:rates}
\end{table}






\pagebreak 
\listoffigures
\listoftables

\end{document}